\documentclass[floatfix,
  aip,
  pop,
  reprint,
  amsmath,amssymb,
]{revtex4-2} 

\usepackage{lmodern}

\usepackage{gensymb}
\usepackage{graphicx}
\usepackage{dcolumn}
\usepackage{bm}
\usepackage{cases}
\usepackage{xcolor}

\usepackage[T1]{fontenc}
\usepackage{newtxtext, newtxmath}
\usepackage{isomath} 

\DeclareMathAlphabet{\mathbfit}{OML}{nxlmi}{bx}{it}

\newcommand{\Alfven}{Alfv\'en }
\newcommand{\para}{\parallel}
\newcommand{\Wci}{\Omega_{ci}}

\newcommand{\Wce}{\Omega_{ce}}

\newcommand{\wpe}{\omega_{pe}}
\newcommand{\Wcj}{\Omega_{cj}}
\newcommand{\wpj}{\omega_{pj}}
\newcommand{\vth}[1]{v_{{\rm th}, {#1}}}
\newcommand{\tbn}{\theta_{\rm Bn}}
\newcommand{\VAe}{V_{\rm A,e}}
\newcommand{\VA}{V_{\rm A}}
\newcommand{\MA}{M_{\rm A}}

\newcommand{\MAHTF}{M_{\rm A}^{\rm HTF}}

\renewcommand{\bm}[1]{\mathbfit{#1}} 
\renewcommand{\eqref}[1]{Eq.~(\ref{#1})} 

\newcommand{\DCW}{DCW } 
\newcommand{\UAW}{UAW } 
\newcommand{\ULW}{ULW } 
\newcommand{\UCW}{UCW } 

\graphicspath{{figure/}}

\usepackage[colorlinks=true, linkcolor=blue, citecolor=blue, urlcolor=blue]{hyperref}

\begin{document}
\setcitestyle{showallnames=false}
\preprint{AIP/123-QED}

\title[Generation of Whistler Waves by Reflected Electrons]{Generation of Whistler Waves by Reflected Electrons and Their Self-Confinement at Quasi-Perpendicular Shocks}

\author{Ruolin Wang}
\email{wangruolin@eps.s.u-tokyo.ac.jp}
\author{Takanobu Amano}
\affiliation{
Department of Earth and Planetary Science, The University of Tokyo, Tokyo 113-0033, Japan.
}

\date{\today}

\begin{abstract}

We investigate the mechanism of whistler-mode wave generation by shock-reflected electrons at quasi-perpendicular collisionless shocks. 
By employing Liouville mapping to construct the electron velocity distribution function in the shock and performing linear instability analysis, we explore whistler wave generation by the mirror-reflected electrons near the upstream edge of the shock transition layer.
We find that the reflected electrons can excite two distinct instabilities with different propagation directions when both the upstream electron beta $\beta_e$ and \Alfven Mach number in the de Hoffmann-Teller frame $\MA/\cos\tbn$ are sufficiently large, where $\MA$ is \Alfven Mach number and $\tbn$ is the angle between the upstream magnetic field and the shock normal. 
In the parameter regime of Earth's bow shock, the instability threshold condition is roughly given by $\MA/\cos\tbn\gtrsim50$.
Since such shocks are super-critical with respect to the whistler critical Mach number, the generated waves cannot propagate upstream and will accumulate in the transition layer. Furthermore, we find that the pitch-angle scattering by the generated waves may trigger secondary instabilities on the same branch. 
We suggest that the sequence of instabilities likely happening within the shock transition layer can efficiently scatter the reflected electrons over a broad range of pitch angles.
Consequently, the reflected electrons may be confined within the shock by the waves generated by themselves. The self-confinement provides the necessary ingredient of stochastic shock drift acceleration, which then offers a plausible mechanism for the electron injection into diffusive shock acceleration.

\end{abstract}

\maketitle

\section{Introduction}
\label{sec:intro}

Collisionless shocks are ubiquitous in space and astrophysical plasmas, examples of which include Earth's bow shock, interplanetary shocks, and supernova remnant (SNR) shocks. The collisionless shocks are widely considered to be major sites of non-thermal, high-energy particle acceleration \cite{blandfordParticleAccelerationAstrophysical1987,bellCosmicrayAccelerationEscape2013}.
The prevailing theoretical framework for this acceleration is diffusive shock acceleration (DSA) \cite{bellAccelerationCosmicRays1978,blandfordParticleAccelerationAstrophysical1978,druryIntroductionTheoryDiffusive1983}. In DSA, particles gain energy by repeatedly crossing the shock front and scattering off plasma waves in both the upstream and downstream regions. 

While remarkably successful in explaining the acceleration of high-energy cosmic rays, DSA faces challenges in efficiently accelerating low-energy particles, particularly electrons. Due to their small gyroradii compared to ions, low-energy electrons do not interact efficiently with long-wavelength magnetohydrodynamic (MHD) turbulence. As a result, it is difficult to account for electron acceleration by DSA unless there is a sufficiently energetic seed population \cite{levinsonElectronInjectionCollisionless1992a,ellisonFirstOrderFermiParticle1990,amanoElectronInjectionHigh2007}. The difficulty is commonly referred to as the electron injection problem.

Observations of radio and X-ray synchrotron emission from young SNRs directly probe the presence of high-energy electrons accelerated at the shocks \cite{koyamaEvidenceShockAcceleration1995,balletXraySynchrotronEmission2006}, which indicates that there exists an efficient electron injection mechanism. 
It is clear that a comprehensive understanding of electron acceleration at collisionless shocks has not yet been established.

While remote observations provide the global context, the relevant microphysical processes can be examined in much greater detail at heliospheric shocks, such as Earth's bow shock. In situ measurements sometimes find that quasi-perpendicular shocks can efficiently accelerate electrons. However, the dependence of acceleration efficiency on shock parameters such as Mach number, plasma beta, and shock obliquity has not yet been completely understood \cite{okaWhistlerCriticalMach2006,trottaElectronAccelerationQuasiperpendicular2019,lindbergElectronAccelerationEarths2024}.

At quasi-perpendicular shocks, shock drift acceleration (SDA) provides an efficient energization mechanism. In the normal incidence frame (NIF), electrons gain energy through the gradient-B drift in the direction anti-parallel to the motional electric field \cite{wuFastFermiProcess1984b,leroyTheoryEnergizationSolar1984}. In the de Hoffman Teller frame (HTF), this process is understood as adiabatic mirror reflection. In the adiabatic picture, however, SDA operates only over a limited residence time of electrons in the shock: after a single reflection or drift cycle, they rapidly escape upstream, preventing further energization.

To reach higher energies, electrons must be confined near the shock front for extended periods. The prolonged residence time allows electrons to undergo sustained drift acceleration, thereby significantly enhancing the energy gain --- a concept central to stochastic shock drift acceleration (SSDA) \cite{amanoObservationalEvidenceStochastic2020,amanoTheoryElectronInjection2022a,katouTheoryStochasticShock2019}. SSDA assumes that particles are subject to strong scattering by intense wave activity in the shock, which makes the residence time longer in a probabilistic manner.

The presence of intense whistler waves has been reported at Earth's bow shock \cite{zhangExtremelyIntenseWhistler1999,hullMultiscaleWhistlerWaves2012,amanoStatisticalAnalysisHighfrequency2024b,okaElectronScatteringHighfrequency2017a}, suggesting that they may play a central role in electron acceleration at the quasi-perpendicular Earth's bow shock.

An important question concerns the origin of these whistler-mode waves. The shock itself can naturally induce anisotropies in the electron velocity distribution function (VDF). For example, the mirror reflection of electrons in the HTF and magnetic-field compression across the shock ramp produce strong pitch-angle anisotropies. These anisotropic distributions provide free energy for kinetic instabilities \cite{tokarPropagationGrowthWhistler1985,garyElectronAnisotropyConstraint2005}.
In particular, kinetic instabilities driven by shock-reflected electrons in the electron foreshock region have been extensively studied using theory and Particle-in-Cell (PIC) simulations \cite{karlickyWavesGeneratedReflected2006,amanoCriticalMachNumber2010a,matsukiyoRELATIVISTICELECTRONSHOCK2011,guoNonthermalElectronAcceleration2014,guoNonthermalElectronAcceleration2014a,bohdanElectronForeshockHighMachnumber2022}.

In this study, we extend the idea and focus on the precise mechanism of whistler wave generation by reflected electrons at quasi-perpendicular shocks. We employ Liouville mapping to construct the VDF with reflected electrons and utilize a semi-analytical method to perform linear stability analysis.
We explore the conditions for whistler wave generation, based on which we discuss the application to electron acceleration.

The rest of this paper is organized as follows. Section \ref{sec:linear} details the method of linear analysis for an arbitrary VDF employed throughout the paper. Section \ref{sec:dist} describes the model of the electron VDF within the shock transition layer based on Liouville mapping.
In Section \ref{sec:result}, we show that the whistler mode waves become unstable at quasi-perpendicular shocks, and investigate the dependence of the instabilities on shock parameters, including the \Alfven Mach number, magnetic obliquity angle, and electron plasma beta. 
We demonstrate, in particular, that a high \Alfven Mach number in HTF is favorable for the wave growth. In Section \ref{sec:discussion}, we suggest that, beyond a threshold HTF \Alfven Mach number, the generated waves may efficiently scatter and confine the reflected electrons, and trigger particle acceleration by SSDA. Finally, the summary and conclusion will be given in Section \ref{sec:conclusion}.

\section{Linear Analysis for Arbitrary Distribution Function}
\label{sec:linear}

We estimate the linear growth or damping rate of a given wave mode using the semi-analytical formalism of \citet{kennelResonantlyUnstableOffangle1967,kennelResonantParticleInstabilities1967}. Fully kinetic dispersion solvers for collisionless plasmas normally require evaluation of a complicated dispersion relation and iterative numerical root finding in the complex frequency plane with well-chosen initial guesses. This is particularly challenging for an arbitrary (or non-Maxwellian) VDF for which an analytical closed form of the dispersion relation is not available \cite{verscharenALPSArbitraryLinear2018}. On the other hand, the semi-analytical method used here can provide an estimate of the growth rate $\gamma$ for a resonant mode without fully solving the kinetic dispersion relation, under the weak-growth approximation $|\gamma / \omega| \ll 1$ where $\omega$ is the real part of the frequency.

The method requires the dispersion relation $\omega = \omega(\bm{k})$ where $\bm{k}$ is the wavenumber vector, and the wave electromagnetic field $\bm{E}$, $\bm{B}$. Given the information, the total growth rate $\gamma$ is obtained by summing the resonance contributions of all plasma species $j$ and all cyclotron harmonics $n$:
\begin{equation}
    \gamma
    = \sum_{j} \sum_{n=-\infty}^{+\infty} \gamma_{j,n},
\end{equation}
where $\gamma_{j,n}$ denotes the contribution of the $n$-th order cyclotron harmonic of species $j$. A convenient general expression for $\gamma_{j,n}$ is given by:
\begin{align}
    \frac{\gamma_{j,n}}{|\omega|} =
    & \frac{\pi}{8 n_{j}} \left| \frac{\omega}{k_\para} \right| \left( \frac{\wpj}{\omega} \right)^2 \notag \\
    & \int_0^\infty d v_\perp v_\perp^2
    \int_{-\infty}^{+\infty} dv_\para \delta \left(v_\para - v_{\mathrm{res},n}\right) \frac{|\psi_{j,n}|^2 \hat{G} f_{j}}{W},
    \label{eq:gamma}
\end{align}
where the quantities are defined as follows:
$k_\para$ and $k_\perp$ are the components of the wavenumber parallel and perpendicular to the background magnetic field;
$n_j$ is the number density;
$\wpj \equiv ( 4 \pi n_j q_j^{2}/m_j )^{1/2}$ is the plasma frequency;
$\Wcj \equiv q_j B/m_j c$ is the cyclotron frequency;
$m_j$ and $q_j$ are the particle mass and charge;
$v_{\mathrm{res}, n} = (\omega - n \Wcj)/k_\para$ is the resonance velocity for the $n$-th cyclotron harmonic;
and $B$ is the background magnetic field.
In addition, the wave energy density $W$ and the weighting function $\psi_{j, n}$ defined below depend on the wave electromagnetic fields:
\begin{align}
  W & \equiv \frac{1}{16\pi}
    \left[
      \bm{B}^* \cdot \bm{B} +
      \bm{E}^* \cdot \frac{\partial}{\partial \omega} (\omega \varepsilon_h) \bm{E}
    \right],
    \\
  \psi_{j,n} & \equiv \frac{1}{\sqrt{2}}
  \left[
    E_{r} J_{n+1}(\rho_j) + E_{l} J_{n-1}(\rho_j)
  \right] +
  \frac{v_\para}{v_\perp} E_{z} J_n(\rho_j),
\end{align}
where $\varepsilon_h$ denotes the Hermitian part of the dielectric tensor, $J_n (\rho_j)$ is the $n$-th order Bessel function with argument $\rho_j \equiv k_\perp v_\perp / \Wcj$, and $E_r = (E_x - iE_y) / \sqrt{2}$, $E_l =(E_x + iE_y) / \sqrt{2}$. Note that the background magnetic field is taken along the $z$ direction.


Finally, the differential operator $\hat{G}$ defined by
\begin{equation}
   \hat{G} \equiv \left( 1 - \frac{k_\para v_\para}{\omega} \right) \frac{\partial}{\partial v_\perp} + \frac{k_\para v_\perp}{\omega} \frac{\partial}{\partial v_\para},
\end{equation}
acts on the gyrotropic VDF $f_j = f_j (v_\para, v_\perp)$ as a function of the parallel ($v_\para$) and perpendicular ($v_\perp$) velocity components with respect to the background magnetic field. It is well known that $\hat{G} f_j$ represents the slope of the VDF along the diffusion curve in the velocity space \cite{gendrinGeneralRelationshipsWave1981}. 
In the present work, we consider only waves with positive energy density (W > 0) in the local plasma rest frame.
It is easily seen that the sign of $\hat{G} f_j$ at the resonance velocity (integrated over $v_\perp$) determines whether the resonance contributes to wave growth or damping. The crucial advantage of this method is that it allows us to evaluate the growth rate for an arbitrary VDF by calculating the derivatives, without the need for finding roots of a nonlinear function.

It is instructive to consider the special case of strictly parallel propagation, i.e., $k_\perp = 0$. Since $\rho_j = 0$ in this case, the Bessel functions reduce to $J_0(\rho_j) = 1$ and $J_n(\rho_j) = 0$ for $n \neq 0$. Therefore, only the resonance $n = 0$ has a nonzero contribution for an electrostatic perturbation with $E_z \neq 0$ and $E_r = E_l = 0$. This is clearly the well-known Landau resonance. Similarly, only the resonance $n = -1$ ($n = +1$) has a nonzero contribution to an electromagnetic perturbation of right-handed (left-handed) polarization with $E_l = E_z = 0$ ($E_r = E_z = 0$). In the general case of oblique propagation $k_\perp \neq 0$, all the cyclotron harmonic resonances potentially contribute to the growth rate.

For the rest of this paper, we consider only the whistler mode with right-handed polarization. Therefore, we will refer to the resonance $n = -1$ as the normal cyclotron resonance. For oblique whistler waves, we consider also the Landau resonance $n = 0$ and the so-called anomalous cyclotron resonance $n = +1$. We ignore other higher-order cyclotron harmonics ($|n| \geq 2$) as the contributions are negligible for the interaction between low-energy electrons and whistler waves. For given $\bm{k}$, we adopt the real frequency $\omega$ and the corresponding electromagnetic field eigenvectors ($\bm{E}$ and $\bm{B}$) obtained by solving the standard cold-plasma dispersion relation \cite{kennelResonantlyUnstableOffangle1967} using the associated dielectric tensor. These quantities are then used in \eqref{eq:gamma}.

\section{Modeling Velocity Distribution Function}
\label{sec:dist}

\subsection{Assumptions}

We model the electron VDF in the shock transition layer by Liouville mapping from prescribed VDFs in the far upstream and downstream regions. Here we use the HTF, in which the motional electric field vanishes and the analysis is considerably simplified \cite{wuFastFermiProcess1984b}.

We assume a one-dimensional steady-state shock with the shock normal along $x$ direction.
The limits $x \rightarrow -\infty$ and $x \rightarrow +\infty$ correspond to the far upstream and downstream regions, respectively, and the shock of finite thickness is centered at $x = 0$. Henceforth, subscripts $1$ and $2$ denote quantities in the far upstream and downstream regions, respectively. The coplanarity plane is taken to be the $x$–$z$ plane, in which both the magnetic field and the flow vectors are contained.

The magnetic field profile across the shock is prescribed as $\bm{B}(x) = \bigl(B_x,\,0,\,B_z(x)\bigr)$, with a constant normal component $B_x = B_1 \cos\tbn$, and a tangential component
\begin{equation}
   B_z(x) =
   \left[
    1 + \frac{r-1}{2}\left(1 + \tanh x\right)
   \right] B_{1} \sin \tbn,
   \label{eq:B_profile}
\end{equation}
where $\tbn$ is the angle between the upstream magnetic field and the shock normal, $B_1$ is the upstream field magnitude, and $r$ is the compression ratio. The upstream tangential component is $B_{z,1} = B_1 \sin\tbn$, and the ratio of the magnitudes of the downstream to the upstream field is $B_2/B_1 = \sqrt{\cos^2 \tbn + r^2 \sin^2 \tbn}$. Note that we denote the magnetic field strength by $B(x) = |\bm{B}(x)|$.

In this study, we assume that the plasma flow in the HTF is everywhere parallel to the local magnetic field, even within the shock transition layer, and denote the parallel flow velocity by $U(x)$. The normal and tangential components of the flow velocity are then determined by condition $U_x/U_z = B_x/B_z$. We see that $U_z$ remains constant across the shock, while $U_x$ deccelerates as the magnetic field compresses. In other words, the frozen-in condition ($U_x B_z = \mathrm{const})$ is satisfied throughout the shock transition. The ratio of downstream to upstream flow speed is given by ${U_{2}}/{U_{1}} = \sqrt{\cos^{2}\tbn / r^2 + \sin^{2}\tbn}$.

It is well established that a finite cross-shock electrostatic potential develops in the shock transition layer due to the inertial difference between the ions and electrons \cite{burgessCollisionlessShocksSpace2015}. In situ observations have revealed a correlation between the magnetic field and potential profiles \cite{hullElectronTemperatureHoffmannTeller2000}, indicating that the potential variation approximately scales as $\delta \Phi \propto |\delta B|$. 
From first principles, this scaling is consistent with the requirement that the potential must account for the reduction of the bulk kinetic energy of the plasma (primarily ions) across the shock. Given that ($U_x \times B_z$) remains relatively constant in the shock frame, it is expected that the potential and the magnetic field are related.
Considering quasi-perpendicular shocks where $\delta B \approx \delta B_z$, we adopt the cross-shock potential profile:
\begin{equation}
  \Phi(x) = \frac{1}{2} \left( 1 + \tanh x \right) \Delta\Phi,
\end{equation}
where $\Delta \Phi = \Phi_2 - \Phi_1$ represents the potential jump between the upstream ($\Phi_1$) and downstream ($\Phi_2$) regions.

Since the potential physically scales with the ion flow kinetic energy in the upstream (as confirmed by observations \cite{hullElectronTemperatureHoffmannTeller2000}), the following normalized potential gives a reasonable measure of the strength of the potential
\begin{equation}
\tilde{\phi} = \frac{ 2 e \Delta \Phi}{m_i V_{1}^2},
\end{equation}
where $e$ is the elementary charge, $m_i$ is the ion mass, and $V_{1} = U_1 \cos \tbn$ is the normal component of the upstream flow speed.

We then assume that the interaction of electrons with the shock is described by the adiabatic theory. In other words, the total energy $ K \equiv m_e v_{\para}^2 / 2 + M B(x) - e \Phi(x)$ and the first adiabatic invariant $M \equiv m_e v_\perp^2/2B(x)$ (where $m_e$ is the electron mass) are both constant. This approximation requires that the gradient scale of the shock is much larger than the electron gyroradius, which is a reasonable assumption for low-energy electrons that we consider here. We should point out that the length scale is implicitly normalized by the shock transition scale size in defining the magnetic field profile \eqref{eq:B_profile}. As long as the transition scale is much larger than the electron gyroradius, only the relative position over the entire shock transition is relevant for the modeling of VDF.

In the present study, we use constant values of $r = 4$ and $\tilde{\phi} = 0.05$. Obviously, $r = 4$ corresponds to the compression ratio for the strong shock limit as predicted by the Rankine-Hugoniot relationships with the polytropic index of $\gamma = 5/3$. 
The choice of $\tilde{\phi} = 0.05$ represents HTF cross shock potential.
Observational studies indicate that the HTF potential is significantly smaller than the NIF potential and tends to decrease at higher Mach numbers (Hanson 2019; Dimmock 2012).
Although slightly smaller than the mean value $\tilde{\phi} \sim 0.1$ estimated by the statistical analysis of Earth's bow shock observations \cite{hullElectronTemperatureHoffmannTeller2000,schwartzElectronHeatingPotential1988}, $\tilde{\phi} = 0.05$ strill remians within the range of statistical variations.

We note that the model shock profile adopted here may not fully represent a realistic shock structure, exhibiting highly dynamic and non-stationary behaviors. For instance, a non-monotonic magnetic field profile with the overshoot and undershoot structure is a common feature of super-critical collisionless shocks \cite{burgessCollisionlessShocksSpace2015}, which, however, introduces additional complexity in the Liouville mapping \cite{hullElectronHeatingPhase2001}. Similarly, the presence of non-stationary behaviors such as shock self-reformation \cite{lembegeNonstationarityTwodimensionalQuasiperpendicular1992,scholerQuasiperpendicularShocksLength2003,hadaShockFrontNonstationarity2003,krasnoselskikhNonstationarityStrongCollisionless2002} and rippling \cite{winskeMagneticFieldDensity1988,lowePropertiesCausesRippling2003,johlanderRippledQuasiperpendicularShock2016}, strictly speaking, violates the steady-state assumption. Nevertheless, we believe that the overall behavior of the electron interaction with the shock, particularly the adiabatic mirror reflection, is captured reasonably well by the simplified model.

\subsection{Liouville Mapping}

Liouville's theorem states that, in the absence of binary particle collisions, the phase-space density is constant along the particle trajectory in phase space \cite{freidbergPlasmaPhysicsFusion2008}. For a steady-state and monotonic electromagnetic field profile considered here, the VDF at any point $x$ in the shock can be related to those defined either in the upstream or downstream. More specifically, the local VDF in HTF under the gyrotropic assumption $F(x, v_{\para}, v_{\perp})$ can be constructed by mapping the velocity coordinate $(v_{\para}, v_{\perp})$ to the source $(v_{\para, j}, v_{\perp, j})$ and evaluating the VDF defined there, where $j = 1$ ($j = 2$) indicates the mapping to the far upstream (downstream) region. The mapping between the local and source velocity coordinates is given by:
\begin{align}
v_{\perp, j}^2 & = v_\perp^2\,\frac{B_j}{B(x)}, 
\\
v_{\para, j}^2 & = v_\para^2
+ \frac{2 M}{m_e} \left( B(x) - B_j \right)
- \frac{2 e}{m_e} \left( \Phi(x)-\Phi_j \right).
\end{align}

The mapping naturally defines the regions of accessibility for particles originating from the upstream and downstream. Clearly, a particle with the phase-space coordinate $(x,v_\para,v_\perp)$ is accessible to the region $j$ only when $v_{\para,j}^2 > 0$. Therefore, the condition $v_{\para,j}^2=0$ provides an accessibility boundary. More specifically, the region accessible to the upstream particles is defined by
\begin{align}
v_{\para, 1}^2 =
v_{\para}^2 +
v_{\perp}^2 \left( 1 - \frac{B_1}{B(x)} \right) -
\frac{2e}{m_e} \left( \Phi(x) - \Phi_1 \right) \geq 0,
\label{eq:upstream_accessible}
\end{align}
and the equality sign indicates an ellipse in $v_{\para}-v_{\perp}$ plane (for $B_1 \leq B(x)$), below which the upstream particles are not accessible. Similarly, the region accessible to the downstream particles is defined by
\begin{align}
v_{\para, 2}^2 =
v_{\para}^2 +
v_{\perp}^2 \left( 1 - \frac{B_2}{B(x)} \right) -
\frac{2e}{m_e} \left( \Phi(x) - \Phi_2 \right) \geq 0.
\label{eq:downstream_accessible}
\end{align}
In this case, the equality sign gives a hyperbola in $v_{\para}-v_{\perp}$ plane (for $B_2 \geq B(x)$), above which the downstream particles are not accessible.

It is clear that the region defined by the intersection of the two regions \eqref{eq:upstream_accessible} and \eqref{eq:downstream_accessible} (i.e., above the ellipse and below the hyperbola) is accessible to both the upstream and downstream particles. Therefore, we have to introduce an additional constraint to define whether the source is the upstream or the downstream in constructing the local VDF. A natural choice is to select the source based on the causality argument. In other words, we assume that only a particle with a positive (negative) parallel velocity in the far upstream (downstream) can reach the shock transition layer. This simply means that,  within the intersection, the positive (negative) parallel velocity region at any point $x$ should be mapped to the upstream (downstream) region. In this way, we can define the Liouville mapping without any ambiguity.

To proceed to the mathematical definition, we denote the upstream accessible region $\mathcal{U} = \{(x, v_{\para}, v_{\perp}) \, | \, v_{\para, 1}^2 > 0\}$, the downstream accessible region $\mathcal{D} = \{(x, v_{\para}, v_{\perp}) \, | \, v_{\para, 2}^2 > 0\}$, the positive parallel velocity region $\mathcal{V}_{+} = \{(x, v_{\para}, v_{\perp}) \, | \, v_{\para} > 0\}$, and the negative parallel velocity region $\mathcal{V}_{-} = \{(x, v_{\para}, v_{\perp}) \, | \, v_{\para} < 0\}$, respectively. With this notation, the Liouville mapping is defined as follows:
\begin{align}
& F_\mathrm{LM} (x, v_{\para}, v_{\perp}) \equiv \nonumber \\
&
\begin{cases}
F_1(v_{\para,1}, v_{\perp,1}),
& \text{if} \,
(x,v_\para,v_\perp)\in \mathcal{D}^c \cup (\mathcal{U}\cap\mathcal{D}\cap\mathcal{V}_{+}), \\
F_2(v_{\para,2}, v_{\perp,2}),
& \text{if} \,
(x,v_\para,v_\perp)\in \mathcal{U}^c \cup (\mathcal{U}\cap\mathcal{D}\cap\mathcal{V}_{-}).
\end{cases}
\label{eq:lm_dist}
\end{align}
where $\mathcal{A}^c$ indicates the complement of $\mathcal{A}$. Note that we have taken the positive sign for $v_{\para, 1}$ and the negative sign for $v_{\para, 2}$, to be consistent with the causality. We prescribe the far upstream ($j=1$) and downstream ($j=2$) electron VDFs as the drifting isotropic Maxwellian specifically for the portion of the distribution representing particles moving toward the shock, given by,
\begin{equation}
F_j(v_\para, v_\perp) = \frac{n_j}{(2\pi \vth{j}^2)^{3/2}}
\exp \left[-\frac{(v_\para - U_j)^2 + v_\perp^2}{2 \vth{j}^2}\right],
\label{eq:fj}
\end{equation}
where $n_j$, $U_j$, and $\vth{j}$ are the density, bulk drift velocity, and thermal velocity, respectively.

The upstream VDF parameters are the inputs to the model. On the other hand, we have to introduce additional assumptions to determine the downstream VDF parameters. We should mention that, in general, if we simply consider Liouville mapping from the upstream electrons, the downstream density and the bulk velocity will not be consistent with the Rankine-Hugoniot relations. This may be easily understood by the fact that a fraction of the incoming electrons are reflected back to the upstream, resulting in a reduction of the downstream density. Similarly, the bulk acceleration by the cross-shock potential will make the bulk velocity inconsistent with the assumed flow profile, which will also affect the density through the number flux conservation.

To determine the downstream parameters, we assume that the upstream electrons adiabatically transmitted through the shock are fully thermalized to form a Maxwellian in the downstream rest frame. Namely, we impose the condition that the downstream bulk velocity is equal to the MHD flow velocity $U_2$. The downstream density is simply equal to the density obtained from the velocity moment of the Liouville-mapped VDF. The downstream thermal velocity is then determined by the energy conservation. More specifically, we first construct the downstream VDF on a regular velocity-space mesh by Liouville mapping the transmitting upstream electrons to the downstream: $F(v_\para, v_\perp) = F_1 (v_{\para,1}, v_{\perp,1})$ for $ (v_\para, v_\perp) \in \mathcal{U} \cap \mathcal{V}_{+}$, otherwise $F(v_\para, v_\perp) = 0$. We then calculate the downstream density $n_2$ and the thermal velocity $\vth{2}$ by taking the following velocity moments by numerical integration:
\begin{align}
    n_2
    &=
      \int_{-\infty}^{\infty} dv_\parallel
      \int_{0}^{\infty} 2\pi v_\perp\, dv_\perp
      F(v_\perp, v_\parallel),
    \\
    \frac{3}{2} n_2 \vth{2}^2
    &=
      \int_{-\infty}^{\infty} dv_\parallel
      \int_{0}^{\infty} 2\pi v_\perp\, dv_\perp
      F(v_\perp, v_\parallel) \varepsilon,
\end{align}
where $\varepsilon = ((v_\para - U_2)^2 + v_\perp^2)/2$ is the particle kinetic energy per unit mass in the downstream rest frame.
Since the electrons transmitted through the shock gain energy mainly from the cross-shock potential, the downstream thermal velocity is roughly given by $m_e \vth{2}^2 \sim e \Delta \Phi$, which is consistent with the observed electron heating at the Earth's bow shock \cite{hullElectronTemperatureHoffmannTeller2000,schwartzElectronHeatingPotential1988}.

To proceed with the concrete model, we have to parameterize the upstream and downstream VDFs in terms of the shock parameters. In addition to the shock obliquity $\tbn$ already introduced, the \Alfven Mach number $\MA = V_1 / \VA$ defined in terms of the normal component of the upstream flow velocity (i.e., in the NIF), and the upstream electron plasma beta $\beta_e = 2 \vth{1}^2/\VAe^2$ are the key shock parameters. Note that the \Alfven velocity $\VA \equiv B_1/\sqrt{4 \pi n_1 m_i}$ (which ignores the small electron contribution to the mass density) and the electron \Alfven velocity $\VAe \equiv B_1/\sqrt{4 \pi n_1 m_e}$ are defined in the far upstream. We also introduce the \Alfven Mach number in HTF by $\MAHTF = U_1 / \VA = \MA / \cos \tbn$, which will frequently appear in the following discussion.

\subsection{Pitch-angle Diffusion}

The Liouville mapping employed here considers a laminar shock transition without any small-scale or high-frequency electromagnetic fluctuations, allowing the use of adiabatic theory for the electron–shock interaction. On the other hand, we shall use the VDF constructed in this way to investigate possible wave generation mechanisms, which will then lead to scattering of electrons and violate the assumption itself. In addition, it should be noted that the accessibility boundary in the Liouville mapping produces an artificial discontinuity in VDF because, in general, there is no reason why the phase space density is continuous across the boundary separating particles originating from different regions. In reality, such a discontinuity in velocity space will likely be smeared out quickly by either pre-existing or self-generated fluctuations.

Ideally, we have to deal with the electron transport in the presence of scattering in a self-consistent manner to resolve the apparent contradiction and smooth the artificial discontinuity. However, this requires much more complicated modeling and is far beyond the scope of this paper. Instead, we take into account the non-adiabatic effect, rather in an ad-hoc manner, by applying the diffusion in pitch-angle to the VDF obtained from the Liouville mapping based on the adiabatic assumption. More specifically, we use the pitch-angle diffusion equation for VDF defined in the local plasma rest frame $f(v, \mu, t)$ 
\begin{equation}
\frac{\partial f}{\partial t}
= \frac{\partial}{\partial \mu}
\left[ D_{\mu\mu}(1-\mu^2)\,\frac{\partial f}{\partial \mu} \right],
\label{eq:paevo}
\end{equation}
where $D_{\mu\mu}$ is the pitch-angle diffusion coefficient, $v$, $\mu$ are the velocity magnitude and the pitch-angle cosine, both defined in the local plasma rest frame. In order to mimic the non-adiabatic pitch-angle scattering, we evolve VDF in time, starting from the Liouville mapping result $F_\mathrm{LM} (v_{\para}, v_{\perp})$ defined by \eqref{eq:lm_dist} transformed into the local plasma rest frame: $f(v, \mu, t = 0) \equiv F_\mathrm{LM} (v \mu + U(x), v \sqrt{1 - \mu^2})$.

For analytical tractability, we assume $D_{\mu\mu}$ is constant. In this case, by expanding VDF in terms of Legendre polynomials $P_m(\mu)$
\begin{align}
f (v,\mu, t)
= \sum_{m=1}^{N} \frac{2 m + 1}{2} P_m(\mu) g_m(v, t),
\label{eq:ffm}
\end{align}
we can write the analytic solution for each expansion coefficient as follows
\begin{align}
g_m (v, t) = g_m (v, 0) \exp \left[ - \frac{1}{2} m (m + 1) D_{\mu\mu} t \right].
\end{align}
The full analytic solution is obtained by substituting this expression back into the expansion. In this way, once the Legendre polynomial expansion is performed for the initial VDF, we can easily obtain the resulting VDF at an arbitrary time $t$. Note that we can easily calculate the expansion coefficients for given $f(v, \mu)$ by evaluating the following integral with Gauss-Legendre quadrature
\begin{align}
g_m (v) = \int_{-1}^{+1} P_m(\mu) f(v, \mu) d \mu
\approx \sum_{i=1}^{N} w_i f(v, \mu_i),
\end{align}
where $\mu_i$ are roots of $P_N(\mu)$ and $w_i = 2 / ((1-\mu_i^2) [P'_N(\mu_i)]^2)$ are the corresponding weights.

It is clear that the normalized time $D_{\mu\mu} t$ quantifies the degree of pitch-angle diffusion: $D_{\mu\mu} t \ll 1$ indicates negligible diffusion, whereas VDF is fully isotropized for $D_{\mu\mu} t \gg 1$. In this study, we consider $D_{\mu\mu} t$ as a free parameter instead of trying to estimate the value of $D_{\mu\mu}$ itself. The idea is that the sensitivity of a possible kinetic instability on $D_{\mu\mu} t$ provides a measure of its persistence against isotropization. If the instability is not too sensitive on $D_{\mu\mu} t$, it may be a robust wave generation mechanism. In the opposite case, the instability may be quickly quenched with only a small amount of pitch-angle scattering. It should be noted that, however, there are a number of reasons why the sensitivity on $D_{\mu\mu} t$ alone does not necessarily quantify its importance. For instance, the instability can always be unstable if the time scale of generating the free energy is faster than $D_{\mu\mu}^{-1}$. Obviously, we need more careful analyses to clarify the role of pitch-angle diffusion on a case-by-case basis. Nevertheless, we believe that the simplified treatment presented here should provide insight into the wave generation mechanisms possibly happening in collisionless shocks.

\subsection{Example}

In the following, we normalize the physical quantities by the far-upstream values; the growth rate and the real frequency by the electron cyclotron frequency $|\Wce|$, and the velocity by the electron \Alfven velocity $\VAe$. Correspondingly, the wavenumber is normalized by $\wpe/c$. With these normalizations, the shock parameters $\MA$, $\tbn$, $\beta_e$ and the ion-to-electron mass ratio $m_i/m_e = 1836$ fully determine the Liouville mapping.

\begin{figure}
\centering
\includegraphics[width=1.0\linewidth]{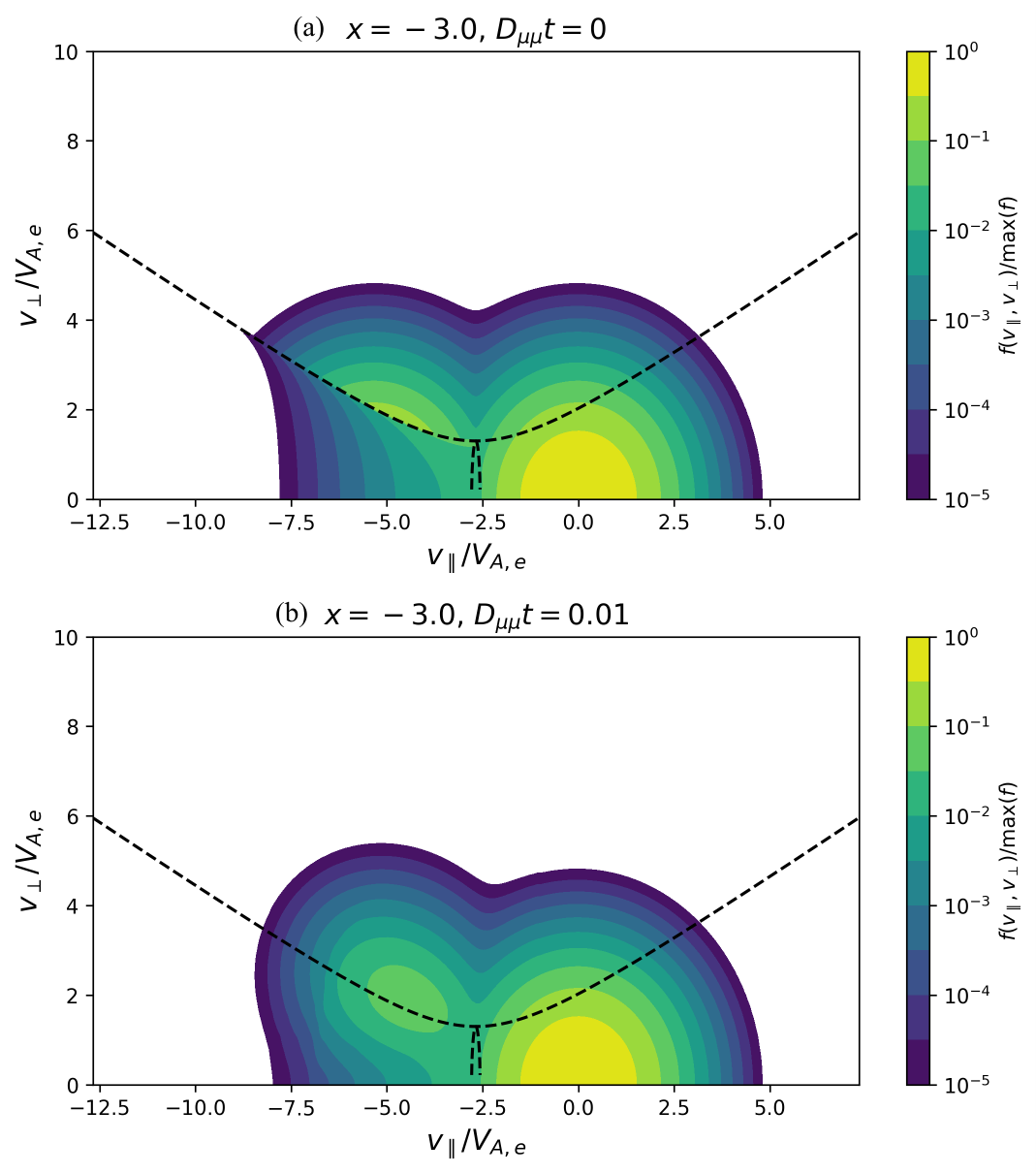}
\caption{\label{fig:diffusion_eff}
Electron VDFs in the plasma rest frame at the upstream edge ($x = -3.0$) for (a) no diffusion ($D_{\mu\mu} t=0$) and (b) weak diffusion ($D_{\mu\mu} t=0.01$). The black dashed lines represent the accessibility boundaries given by Eq.(\ref{eq:upstream_accessible}) (the lower ellipse) and Eq. (\ref{eq:downstream_accessible}) (the upper hyperbola).} 
\end{figure}

An example of the Liouville-mapping and pitch-angle diffusion is shown in Fig.~\ref{fig:diffusion_eff}. We use parameters typical of high-Mach number quasi-perpendicular Earth's bow shock: $\MA = 10, \beta_e = 2.0, \tbn=85^\degree$. In this paper, we will consider, for simplicity, kinetic instabilities driven by the shock-reflected electrons near the upstream edge of the shock transition layer. For this purpose, we show VDFs in the plasma rest frame at a position of $x = -3$ in Fig.~\ref{fig:diffusion_eff} for $D_{\mu\mu} t = 0$ (panel a) and $D_{\mu\mu} t = 0.01$ (panel b), respectively. The color contours show VDFs in a logarithmic scale. The black dashed lines represent the accessibility boundaries given by \eqref{eq:upstream_accessible} (the lower ellipse) and \eqref{eq:downstream_accessible} (the upper hyperbola). The shock speed in HTF given by $-U_1/\VAe \approx -2.68$ can be identified by the center of the ellipse.

There is a clear discontinuous jump in the phase space density across the accessibility boundary for the case $D_{\mu\mu} t = 0$. 
Particles with $v_{\para} < -U_1$ above the hyperbola are those reflected by the shock, while those below the hyperbola are thermal leakage from the downstream. 
A small amount of diffusion $D_{\mu\mu} t = 0.01$ is sufficient to smooth out the discontinuity, while the loss-cone feature persists. Note that the downstream thermal leakage population has often been ignored in modeling the shock-reflected electron populations \cite{amanoCriticalMachNumber2010a}. 
The density of the leakage population depends on the downstream temperature; our model indicates that the density is not large enough to fill the loss cone produced by the adiabatic reflection, which may provide the free energy source for kinetic instability. 
This is a consequence of the assumption that the temperature is determined by the thermalization of the transmitted electrons energized by the cross-shock potential. The majority of the downstream thermal electrons do not have enough energy to overcome the potential barrier to escape back to the upstream.
However, if the downstream temperature were treated as a higher free parameter, the increased leakage could potentially diminish the velocity-space gradient driving the cyclotron-driven modes.

\section{Results}
\label{sec:result}

\subsection{Whistler Wave Self-generation}

\begin{figure}[!tbp]
\centering
\includegraphics[width=\linewidth,height=0.85\textheight,keepaspectratio]{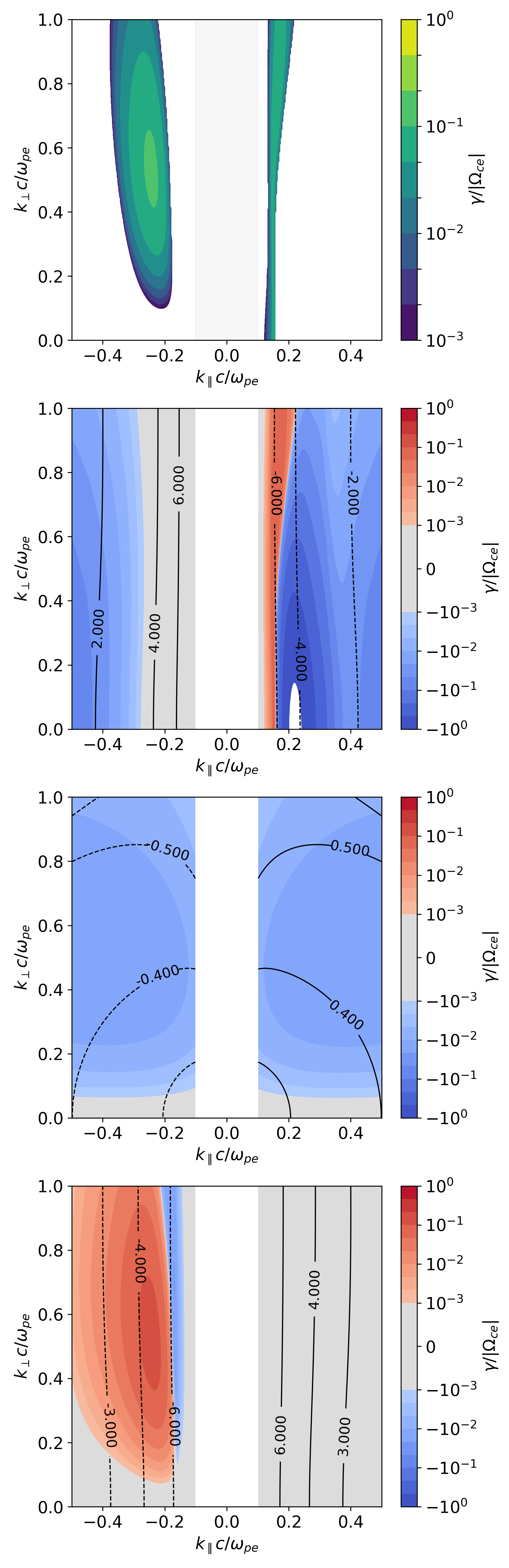}
\caption{\label{fig:fig2}
Linear growth rates in the $(k_{\parallel}, k_{\perp})$ plane driven by the electron VDF with $D_{\mu\mu} t=0.01$ (corresponding to Fig.~\ref{fig:diffusion_eff}b).
From top to bottom, the panels show: the total growth rate, and the individual contributions from the normal cyclotron ($n=-1$), Landau ($n=0$), and anomalous cyclotron ($n=+1$) resonances. The black contours indicate resonance velocities $v_{\mathrm{res},n}$, where the solid and dashed lines represent positive and negative values, respectively.}
\end{figure}

Figure~\ref{fig:fig2} shows the result of growth rate for the weak diffusion case ($D_{\mu\mu} t = 0.01$; see Fig.~\ref{fig:diffusion_eff} (b)), calculated using the semi-analytic method presented in Section \ref{sec:linear}. 
We used the whistler-mode branch of the cold plasma dispersion relation with $\Wce/\wpe = -0.01$, which is typical of the solar wind at 1 au, to calculate the real part and the eigenvector. 
In all the panels, the horizontal and vertical axes represent the wavenumbers parallel ($k_{\para}$) and perpendicular ($k_{\perp}$) to the local magnetic field direction, respectively. A mask is applied to the region $|k_\parallel| < 0.1$ (light shaded gray in the top panel), as the cold plasma approximation becomes unreliable for nearly perpendicular propagation. This region is, however, outside the primary focus of the current study.
Since we choose the magnetic field pointing toward the downstream direction, a positive (negative) $k_{\para}$ indicates that the wave propagation direction is toward downstream (upstream). However, we have to keep in mind that the wave propagation direction represented by $k_{\para}$ is the one with respect to the local plasma rest frame. Since the plasma itself has a finite flow velocity, a proper Doppler shift has to be taken into account to consider the wave propagation with respect to the shock. A more detailed discussion in relation to particle acceleration will be given in Section~\ref{sec:discussion}.

The top panel of Fig.~\ref{fig:fig2} shows the total growth rate $\gamma$, while the second, third, and fourth panels from the top show the contributions from the normal cyclotron ($n = -1$), Landau ($n = 0$), and anomalous cyclotron ($n = +1$) resonances. The total growth rate is shown only for $\gamma/\Wce \geq 10^{-3}$. On the other hand, both positive (in reddish color) and negative (in bluish color) contributions are shown in the bottom three panels to better understand how each resonance contributes to the total growth rate. 
In addition, the resonance velocities for each resonance are shown in solid (for positive) and dashed (for negative) black lines, which enables us to identify the instability-driving velocity-space feature by examining the VDF.
Note that we have ignored the ion contribution, and the total growth rate is calculated by taking the sum of the three resonances (i.e., $\gamma = \gamma_{e,-1} + \gamma_{e,0} + \gamma_{e,+1}$). We have confirmed that the three resonances provide the dominant contributions for the relatively high-frequency ($\omega > \Wci$) whistler waves of our interest here.

It is clear from the top panel of Fig.~\ref{fig:fig2} that there exist two distinct unstable regions in wavenumber space: the one with $k_{\para} > 0$, and the other with $k_{\para} < 0$. Let us first discuss the instability with $k_{\para} > 0$. From Fig.~\ref{fig:fig2}, we see that the positive contribution to the growth rate comes from the normal cyclotron resonance $n = -1$ and the resonance velocity is roughly $v_{\mathrm{res}, -1} \approx -6 \VAe$. Looking at the VDF near the resonance velocity shown in Fig.~\ref{fig:diffusion_eff}, we find that the instability is driven by the loss-cone-like feature of the shock-reflected electron beam. Note that it is well known that the cyclotron resonance occurs when the electron and the whistler wave are counter-streaming with each other.

\begin{figure}[!tbp]
\centering
\includegraphics[width=1.0\linewidth]{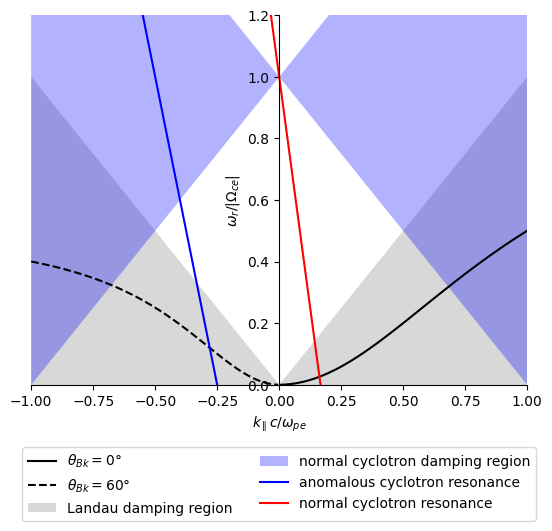}
\caption{\label{fig:fig4} 
$\omega-k$ diagram illustrating the conditions for wave growth and damping. The black curve represents the whistler-mode dispersion relation for propagation angles $\theta_{Bk}=0^\circ$ (solid) and $\theta_{Bk}=60^\circ$ (dashed). 
The shaded areas indicate regions of strong damping by the background electrons: gray for Landau damping ($n=0$) and blue for normal cyclotron damping ($n=-1$). The red and blue lines represent the resonance conditions for beam electrons with $n=-1$ (normal) and $n=+1$ (anomalous), respectively. 
}
\end{figure}

\begin{figure}[!tbp]
\centering
\includegraphics[width=1.0\linewidth]{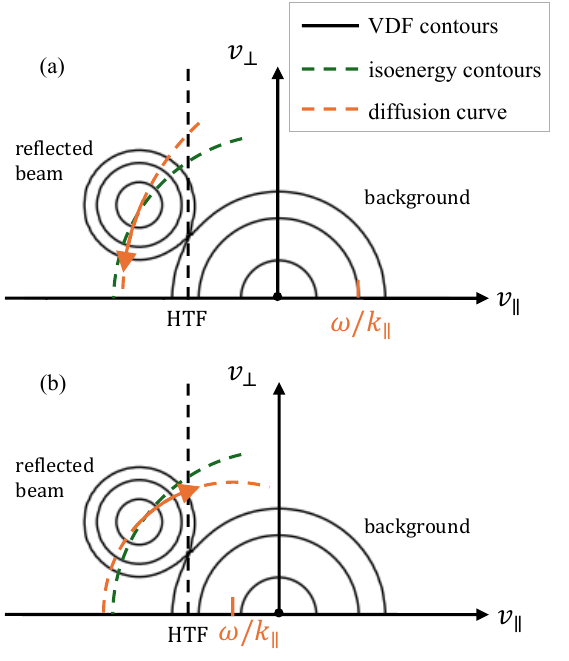}
\caption{\label{fig:fig3} Schematic diagrams illustrating the wave generation mechanism driven by a reflected electron beam in the plasma rest frame for (a) the downstream-directed cyclotron-driven whistler (DCW) instability and (b) the upstream-directed anomalous-driven whistler (UAW) instability. 
The black curves denote contours of the electron VDF, the green dashed curves indicate iso-energy contours in the plasma frame, and the orange dashed curve represents the diffusion path imposed by resonant wave–particle interactions. The orange arrow marks the preferred direction of electron diffusion driven by the local gradient of the VDF. Note that the vertical black dashed lines indicate the shock speed in HTF. In both panels, electrons diffusing inward (inside the isoenergy contours) lose energy to the wave, thereby driving the instability.}
\end{figure}

To better understand the instability mechanism, it is instructive to look at the $\omega-k$ diagram shown in Fig.~\ref{fig:fig4}. The resonance condition is shown in red for a negative parallel velocity, which intersects the whistler-mode branch (black curve). It is known that the intersection can be unstable only when the loss-cone-like structure exists at the resonance velocity \citep{amanoCriticalMachNumber2010a}. The physical mechanism is illustrated schematically in Fig.~\ref{fig:fig3} (a), in which the reflected electrons are represented by a Maxwellian-like beam distinct from the thermal core. Since the beam electrons should be scattered along the diffusion curve defined for a wave with a positive phase velocity $\omega / k_{\para} > 0$, the particles lose energy in the plasma rest frame when they are scattered toward the negative (or upstream) direction. Clearly, this process requires a loss-cone-like gradient in the VDF. The energy lost by the particles is transferred to the wave, which thus drives the instability.

To achieve wave growth, the instability has to overcome damping arising from other resonances. Fig.~\ref{fig:fig4} also illustrates the region in $\omega-k$ space where the damping becomes significant. The effects of Landau damping ($n = 0$) and the normal cyclotron damping ($n = -1$) by the background electrons are significant in the gray-shaded and blue-shaded areas, respectively. 
The damping areas can be roughly estimated by $- k \vth{e} + n \Wce \lesssim \omega \lesssim + k \vth{e} + n \Wce$ with $\vth{e}$ being the electron thermal velocity. The simplest case is perhaps the strictly parallel propagation along the ambient magnetic field ($\theta_{Bk} = 0^\circ$), in which the Landau damping disappears (see Section \ref{sec:linear}). In this case, the loss-cone-beam velocity has to be faster than the thermal velocity of the background electrons to avoid the normal cyclotron damping region \citep{amanoCriticalMachNumber2010a}. For oblique propagation ($k_{\perp} \neq 0$), the Landau damping effect should also be taken into account (see the third panel in Fig.~\ref{fig:fig2}). Therefore, although the result shown in Fig.~\ref{fig:fig2} exhibits a significant spread in $k_{\perp}$, the oblique mode tends to suffer more significant damping in general. Since the wave is destabilized by the normal cyclotron resonance and propagates in the downstream direction, we will refer to this instability as the \DCW (downstream-directed cyclotron-driven whistler) instability in the following.

We now consider the unstable region in $k_{\para} < 0$. This instability is intrinsically oblique with respect to the background magnetic field because, as seen in the last panel of Fig.~\ref{fig:fig2}, it is driven by the anomalous cyclotron resonance $n = +1$ that is absent in a strictly parallel propagation (see Section \ref{sec:linear}). The corresponding resonance velocity $v_{\mathrm{res}, +1} \approx -4 \VAe$ indicates that the mode is also destabilized by the electron beam. The $\omega-k$ diagram in Fig.~\ref{fig:fig4} shows that the resonance condition for the anomalous cyclotron resonance (the blue solid line) may intersect the oblique whistler mode (with $\theta_{Bk} = 60^\circ$) with a negative phase velocity (the black dashed line).

In contrast to the normal cyclotron resonance, both the resonant electron and the whistler wave propagate in the same direction for the anomalous cyclotron resonance case. Therefore, the direction of scattering along the diffusion curve has to be opposite for the resonant particles to lose energy, as illustrated in panel (b) of Fig.~\ref{fig:fig3}. In other words, even though both instabilities are driven by the beam electrons, the instability-driving velocity-space features are different between them. In the following, it is referred to as the \UAW (upstream-directed anomalous-driven whistler) instability, which is again taken from the dominant resonance and the propagation direction. Note that this instability also competes with kinetic damping by the background electrons. We will show in Section \ref{sec:obliquity} that the Landau damping typically provides the primary negative contribution as seen in Fig.~\ref{fig:fig2} (see, also the gray-shaded area in Fig.~\ref{fig:fig4}).

We should note that both of these instabilities were discussed previously in the context of electron acceleration at collisionless shocks. The \DCW instability was proposed as a possible self-generation mechanism of whistler waves by a loss-cone electron beam produced by the adiabatic reflection, but with a much simpler VDF model \cite{amanoCriticalMachNumber2010a}. The oblique whistler generation by the UAW instability is essentially the same as the one discussed by \cite{levinsonElectronInjectionCollisionless1992a,levinsonInjectionElectronsOblique1996} note{.}, who nevertheless considered the wave generation by accelerated electrons under the diffusion approximation as opposed to the coherent reflection by the shock considered here. Note that both require very similar instability conditions in terms of the shock parameters \cite{amanoNonthermalElectronAcceleration2022a}. Nevertheless, their relative importance has not been understood so far. Since our model is able to reproduce both at the same time, we can investigate the roles of these instabilities on equal footing.

\subsection{Effect of Pitch-angle Diffusion}

In general, a linear resonant instability associated with a velocity space gradient tends to smear out the gradient as the wave begins to grow. The system may then approach a marginal stability state, in which the growth rate becomes negligible. Such an evolution can be modeled self-consistently using quasi-linear theory, which, however, requires far more complicated analyses. Instead, we investigate the dependence of the linear growth rate on the assumed degree of pitch-angle diffusion (quantified by $D_{\mu\mu} t$) to infer the robustness of the instability.

\begin{figure*}
\includegraphics[width=0.8\linewidth]{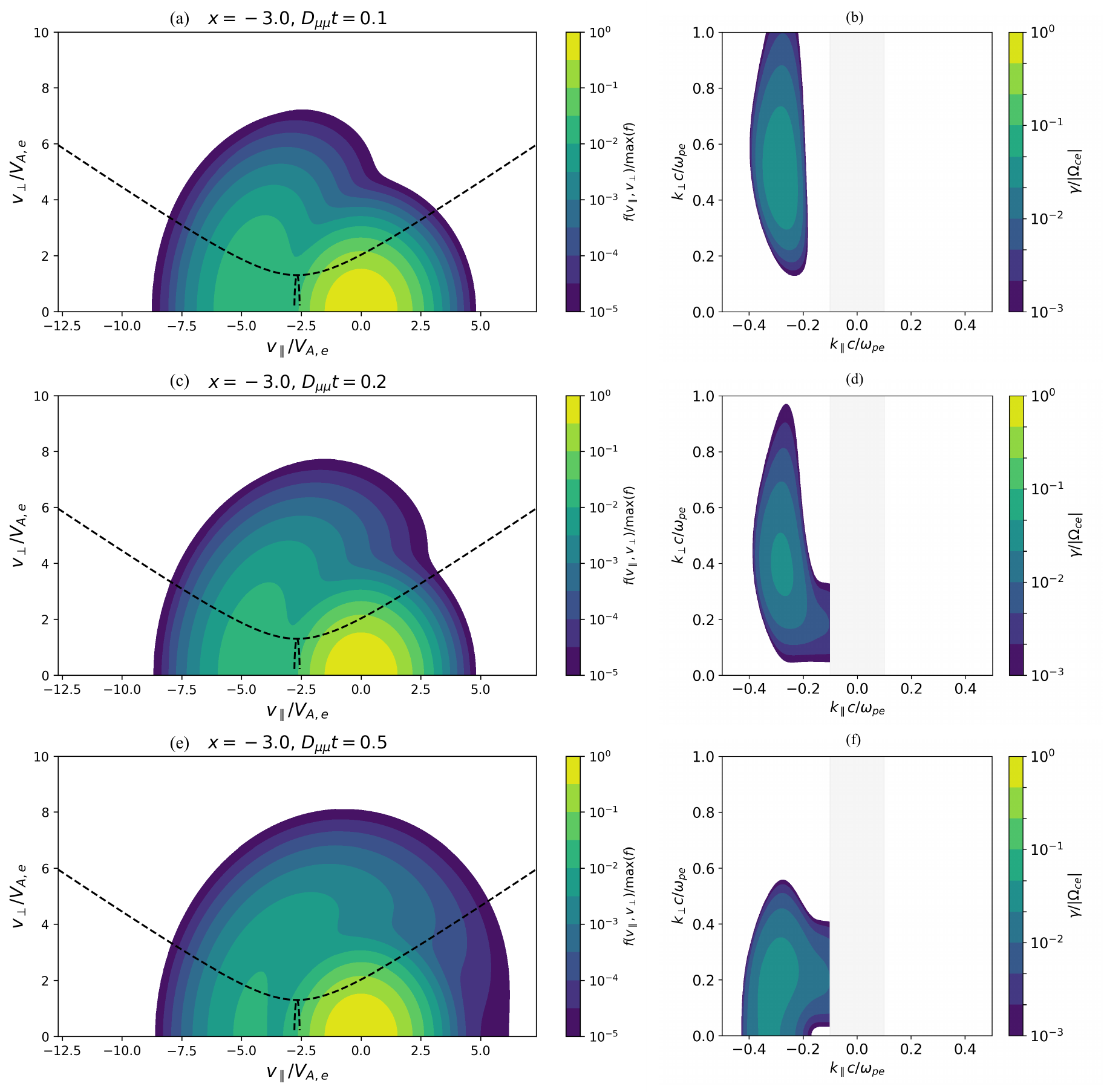}
\caption{\label{fig:fig5} Evolution of the electron VDF and resulting instabilities for increasing diffusion strength. The left panels (a, c, e) show the electron VDFs, while the right panels (b,d,f) show the corresponding growth rate maps. From top to bottom, the degree of diffusion is increased: $D_{\mu\mu} t=0.05$ (top panel), $D_{\mu\mu} t=0.2$ (middle panel) and $D_{\mu\mu} t=0.5$ (bottom panel). }
\end{figure*}

Recall that we have discussed the growth rate for the example of VDF in Fig.~\ref{fig:diffusion_eff} (b) with $D_{\mu\mu} t = 0.01$, i.e., with only very weak diffusion. We here continue to discuss how the different values of $D_{\mu\mu} t$ affect the instability. Figure~\ref{fig:fig5} illustrates the VDFs (left column) and the corresponding growth rates (right column) for three values of $D_{\mu\mu} t = 0.05, 0.2, 0.5$. In the top row, we see that even a modest value of $D_{\mu\mu} t = 0.05$ completely stabilizes the \DCW instability. On the other hand, the \UAW instability persists, albeit with a reduced growth rate. This may be understood intuitively by the fact that the \DCW instability is driven by the loss-cone feature, which is easily smeared out by the pitch-angle diffusion, while the distinct beam feature driving the \UAW instability is still clearly visible in the VDF. As the diffusion proceeds further to $D_{\mu\mu} t = 0.2$ (the middle row), the growth rate of the oblique whistler decreases even further, although it still maintains a positive value.

Interestingly, the pitch-angle diffusion is not necessarily a purely stabilizing process. It smooths the instability-driving VDF gradient and transports electrons to different pitch angles. The diffusing electrons may reach a region in velocity space where they interact with a different wave with a different resonance velocity. The VDF in the middle row of Fig.~\ref{fig:fig5} shows that the diffusing electrons reach a small parallel velocity (nearly stationary in the plasma rest frame), which allows them to satisfy the Landau resonance condition ($n = 0$) with the oblique whistler; the wave on the same branch but with a lower frequency. We have confirmed that the appearance of a very weak but positive growth rate around $k_{\para} c/\wpe \simeq -0.2$, $k_{\perp} c/\wpe \simeq 0.2$ in Fig.~\ref{fig:fig5} (panel d) is due to the Landau resonance contribution (see, Appendix \ref{appendix} for detail). As the diffusion proceeds even further to $D_{\mu\mu} t = 0.5$, the diffusing electrons reach positive parallel velocity as seen in Fig.~\ref{fig:fig5} (panel e). The structure of VDF at this stage has a loss-cone-like feature in the positive parallel velocity region, i.e., on the opposite side of the initial loss-cone beam. This is the reason why there appears a positive growth rate in anti-parallel propagation $k_{\para} c/\wpe \simeq -0.3$, $k_{\perp} c/\wpe \simeq 0$ in Fig.~\ref{fig:fig5} (panel f). In Appendix \ref{appendix}, we show that the contribution to the growth rate at anti-parallel propagation comes from the normal cyclotron resonance ($n = -1$), with the resonance velocity around the newly-appeared loss cone. It should be noted that the contributions from the anomalous cyclotron resonance $(n = +1)$ and the Landau resonance ($n = 0$) both remain positive at this stage.

In summary, the reflected electron beam drives two primary instabilities via two different resonances: the normal cyclotron resonance for the \DCW instability and the anomalous cyclotron resonance for the \UAW instability. As the beam is scattered and diffuse in pitch angle by the primary instabilities (in particular by the \UAW instability), a sequence of secondary instabilities may also arise. We refer to the first one as the \ULW (upstream-directed Landau-driven whistler) instability, since it becomes unstable when the beam electrons decelerate sufficiently to satisfy the Landau resonance with the oblique whistler wave. The next one, which we call the \UCW (upstream-directed cyclotron-driven whistler) instability, becomes unstable when the beam electrons diffuse even further and then satisfy the normal cyclotron resonance with the wave on the same branch, but propagating nearly along the ambient magnetic field. The four instabilities are summarized in Table~\ref{tab:whistler_naming}.

The existence of a single instability may result in scattering of particles only in a limited region in velocity space. However, the sequence of instabilities implies that the scattering may happen over nearly the entire pitch-angle once the primary instabilities are driven unstable. This will naturally lead to isotropization, which is indeed a necessary ingredient of particle acceleeration by SSDA. Therefore, we think that understanding the conditions to excite the primary instabilities that trigger the secondary instabilities is crucial. In particular, the \UAW instability plays the key role in that it is more robust, and the resulting pitch-angle scattering tends to form a VDF favorable for the secondary instabilities.

\begin{table}[tbp]
\centering
\caption{Summary of instabilities. Acronyms denote the propagation direction (D: downstream, U: upstream) and the dominant resonance type (C: cyclotron, A: anomalous, L: Landau). The wave propagation direction is defined with repect to the local plasma rest frame.}
\label{tab:whistler_naming}
\begin{tabular}{lll}
\hline
Instability & Propagation direction & Dominant resonance \\
\hline
DCW & Downstream & Normal cyclotron ($n=-1$) \\
UAW & Upstream & Anomalous cyclotron ($n=+1$) \\
ULW & Upstream & Landau ($n=0$) \\
UCW & Upstream & Normal cyclotron ($n=-1$) \\
\hline
\end{tabular}
\end{table}

\subsection{Obliquity Dependence}
\label{sec:obliquity}

The generation of oblique whistler waves by an electron beam (or heat flux) via the anomalous cyclotron resonance has been discussed in various contexts, including the solar wind, solar flares, and intercluster medium \cite{vaskoWhistlerFanInstability2019,roberg-clarkWaveGenerationHeat2018,roberg-clarkScatteringEnergeticElectrons2019a,verscharenSelfinducedScatteringStrahl2019, fromentWhistlerWavesGenerated2023a,jeongStabilityElectronStrahl2022}. In the simplest case of a Maxwellian beam parallel to the background magnetic field, the necessary condition for the instability is that the beam velocity exceeds the wave phase velocity, which is approximately given by $\VAe/2$ for the whistler wave \cite{verscharenSelfinducedScatteringStrahl2019}. It is essentially the condition illustrated in panel (b) of Fig.~\ref{fig:fig3}, meaning that the instability mechanism is the same as the instability (UAW) discussed here, although the detailed shape of the VDF is different.

If we consider the adiabatic reflection of an electron with the initial velocity equal to the upstream flow velocity, we can estimate the beam velocity $v_b$ in the upstream rest frame roughly as twice the shock velocity in the HTF:
\begin{align}
\frac{v_b}{\VA} \sim 2 \MAHTF
\label{eq:vr}
\end{align}
Accordingly, the instability condition $v_b \gtrsim \VAe/2$ can be rewritten as
\begin{align}
\MAHTF \gtrsim \frac{\alpha}{4} \sqrt{\frac{m_i}{m_e}},
\label{eq:ma_htf_cond}
\end{align}
where we have introduced a numerical factor of order unity $\alpha$, which is dependent on the detailed structure in VDF. While it does not provide the exact threshold, it is clear that a high HTF Mach number is favorable for the instability.

In reality, the instability can grow only if the beam contribution (via the anomalous cyclotron resonance) overcomes both Landau and normal cyclotron damping by the background thermal electrons. The competition between the growth and damping from different resonances ultimately determines the instability condition. It is clear from Fig.~\ref{fig:fig4} that a higher beam velocity makes the resonant wavenumber smaller and the damping by the normal cyclotron damping weaker. On the other hand, the Landau damping is always present for a moderate electron plasma beta $\beta_e \gtrsim 1$. For a Maxwellian beam, it is possible to derive an analytical expression for the instability condition taking into account the anomalous cyclotron and Landau resonances for a given density and a temperature of the beam \cite{verscharenSelfinducedScatteringStrahl2019}. On the other hand, the VDF in our model is far from Maxwellian and is dependent on the various shock parameters, including $\MAHTF$ and $\beta_e$.

\begin{figure*}
\centering
\includegraphics[width=0.8\linewidth]{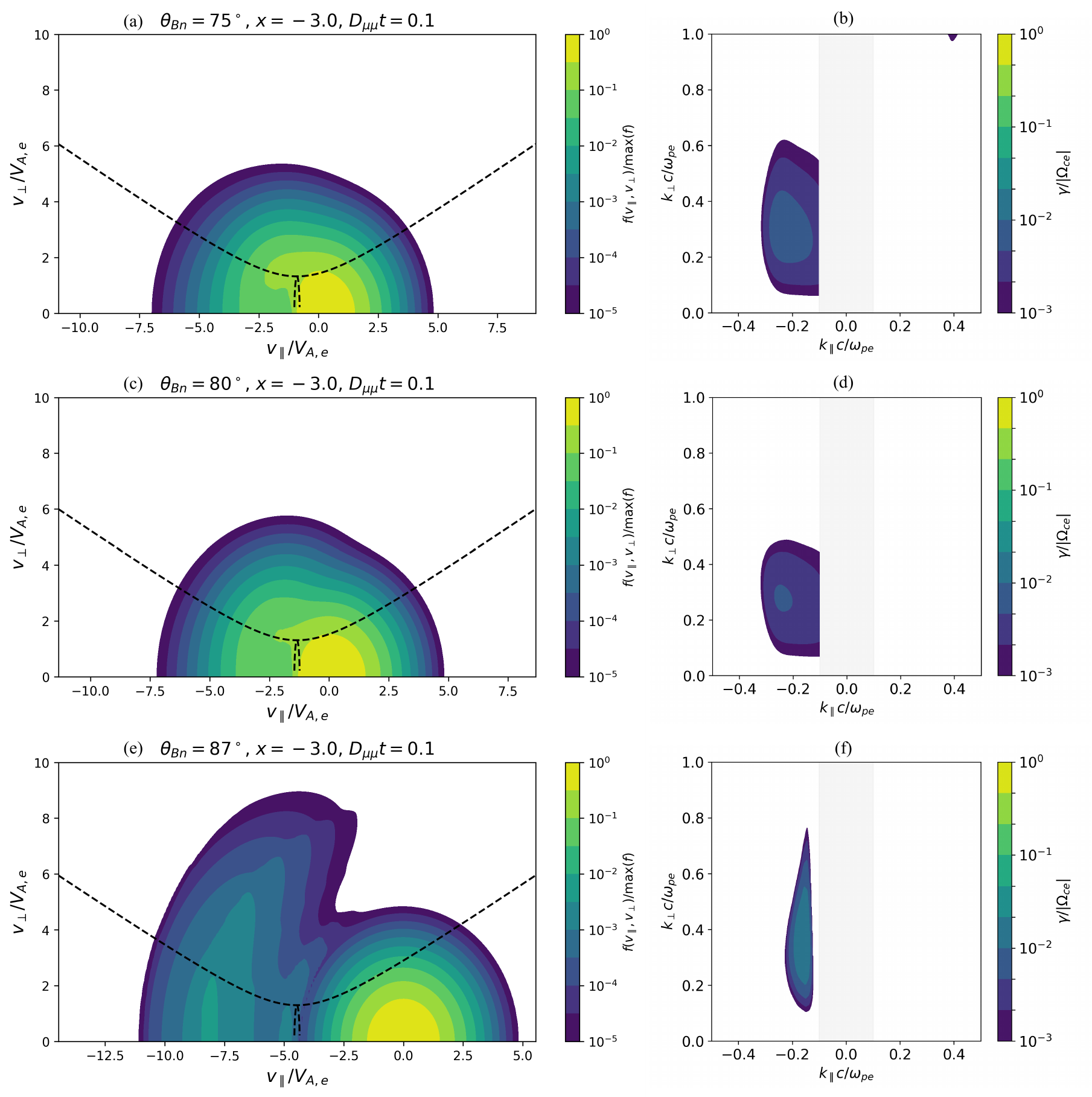}
\caption{\label{fig:fig6} Dependence of the electron VDF and resulting instabilities on the shock obliquity, \(\theta_{Bn}\) with fixed diffusion parameter $D_{\mu\mu} t=0.1$. The format is the same as in Fig.~\ref{fig:fig5}. From top to bottom, the shock obliquity increases: $\tbn = 75 \degree, 80 \degree, 87 \degree$
}
\end{figure*}

Given the theoretical expectation that high-obliquity shocks are favorable for the \UAW instability, we examine the dependence on shock obliquity $\tbn$ while keeping the \Alfven Mach number $\MA$ (typical of Earth's bow shock) constant. We use a fixed value of $D_{\mu\mu} t = 0.1$ corresponding to moderate pitch-angle diffusion that smooths out the loss cone while preserving the beam structure. Figure~\ref{fig:fig6} shows the VDFs (left column) and the corresponding growth rates (right column) for three values of shock obliquity $(\tbn = 75 \degree, 80 \degree, 87 \degree)$.

At a lower obliquity angle $\tbn = 75 \degree$ (the top row), the reflected electrons do not form a distinct beam. Nevertheless, an unstable region exists at oblique propagation $k_{\para}c/\wpe \sim -0.2$, $k_{\perp}c/\wpe \sim 0.3$. We have confirmed (not shown) that the instability is driven by the Landau resonance rather than the anamalous cyclotron resonance. Therefore, it is essentially the same as the \ULW instability, but in this case, it is a primary instability. The reflected electron population progressively separates from the thermal core as $\tbn$ increases, which makes the anomalous cyclotron resonance more important. We have confirmed that the contributions of the \UAW and the \ULW instabilities become comparable at $\tbn = 80 \degree$ for the growth rate (the middle row). As previously demonstrated in Fig.~\ref{fig:fig5} (a), the \UAW instability dominates as the shock obliquity increases to $\tbn = 85 \degree$.

In order to realize efficient pitch-angle scatterings within the shock transition layer, the growth rate must be sufficiently high. If we consider the typical convection time of the upstream plasma through the ion-gyroradius scale transition layer $\sim \Wci^{-1}$, we can estimate that the instability growth rate should roughly be $\gamma/\Wci \gtrsim 10$, or equivalently, $\gamma/\Wce \gtrsim 0.01$. While the maximum growth rate for the case of $\tbn = 85 \degree$ satisfies the condition, the cases of $\tbn = 75 \degree$ and $80 \degree$ appear to be marginal. Therefore, we may roughly estimate these marginal cases to correspond to the critical condition. This indicates that $\alpha \sim 5$ in \eqref{eq:ma_htf_cond}, or $\MAHTF \gtrsim 50$ gives the instability condition for the present set of parameters.

However, the relationship between shock obliquity and the instability growth rate is not monotonic. At a very high obliquity $\tbn = 87 \degree$ (the bottom row), the growth rate decreases again. It is readily seen from the VDF that the reduction in the growth rate comes from the smaller density of the reflected electrons. It is well known that the reflection efficiency rapidly decreases as obliquity increases because the number of upstream electrons outside the loss cone becomes smaller \cite{wuFastFermiProcess1984b,leroyTheoryEnergizationSolar1984}. This can obviously be partially compensated for by an increase in the upstream electron temperature, which will be investigated in the next subsection.

\subsection{Plasma Beta Dependence}

Under the assumption of adiabatic reflection, the number density of the reflected electron $n_b$ can be easily calculated by integrating the far upstream electron VDF over the region outside the loss cone:
\begin{align}
n_b &=
\int_{0}^{\infty} dv_{\para}
\int_{v_{\perp, \mathrm{min}}}^{\infty} 2 \pi v_{\perp} dv_{\perp}
F_1(v_{\para}, v_{\perp})
\end{align}
where $v_{\perp, \mathrm{min}} = \sqrt{(v_{\para}^2 + (2 e/m_e) \Delta \Phi)/(B_2/B_1 - 1)}$. At a nearly perpendicular shock, the majority of electrons are inside the loss cone, and the cross-shock potential effect becomes negligible. This allows us to estimate the dependence of $n_b$ on the shock parameter. Adopting the approximation $v_{\perp, \mathrm{min}} \sim \MAHTF \VA / \sqrt{B_2/B_1 - 1}$ and assuming that the density is dominated by the electrons near the loss-cone boundary, we obtain
\begin{align}
n_b &\propto
\exp\left(- \frac{(\MAHTF)^2}{(B_2/B_1-1) \beta_e} \frac{m_e}{m_i} \right),
\end{align}
which indicates that a higher electron plasma beta $\beta_e$ increases the reflection efficiency for a given value of $\MAHTF$.

\begin{figure*}
\centering
\includegraphics[width=0.8\linewidth]{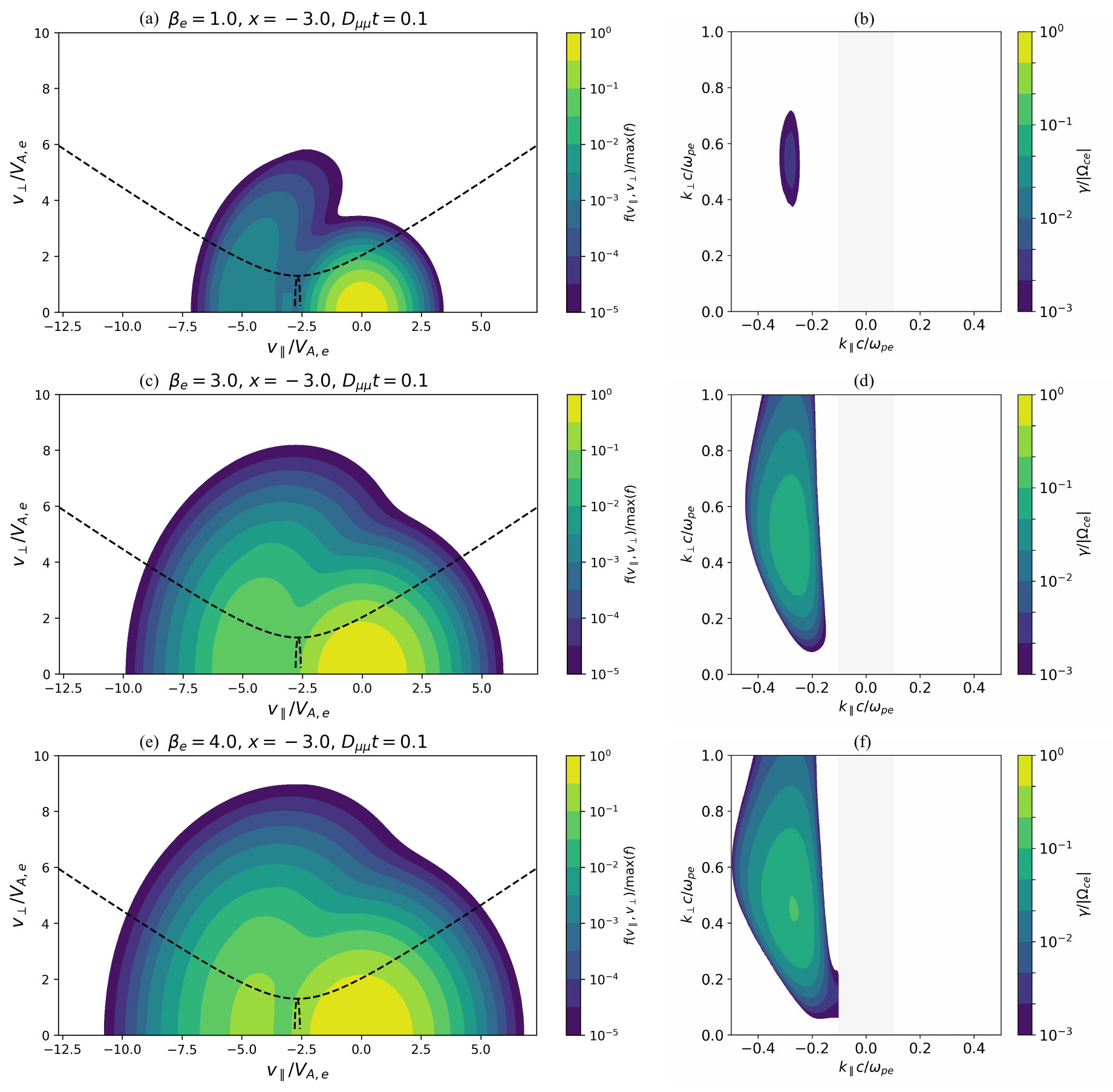}
\caption{\label{fig:fig7}  Dependence of the electron VDF and resulting instabilities on the upstream electron plasma beta (\(\beta_e\)). The format is same as in Fig.~\ref{fig:fig5}. From top to bottom, the electron beta increases: $\beta_e=1.0$, $3.0$ and $4.0$.}
\end{figure*}

The electron plasma beta dependence in Fig.~\ref{fig:fig7} confirms the expected trend. In Figure \ref{fig:fig7}, the VDFs (left column) and the corresponding growth rates (right column) are shown for four values of the electron betas $\beta_e=1.0, 3.0, 4.0$, while the other parameters remain fixed: $\MA=10$, $\tbn=85^\degree$, $D_{\mu\mu} t=0.1$. It is clear that the beam density increases with $\beta_e$, and the instability growth rate also increases accordingly. We have confirmed that the growth rate is always dominated by the \UAW instability for higher $\beta_e$ cases.

Note that the increased electron temperature should also enhance the Landau damping by the background electrons, which in turn tends to suppress the instability. The results shown in Fig.~\ref{fig:fig7} indicate that the increased instability drive overcomes the increased Landau damping. This may be understood by looking at the dependence of the Landau damping rate: $\gamma_{e,0} \propto -\left.\partial f/\partial v \right|_{\omega/k_\para}$, where the phase velocity is on the order of the elecron \Alfven velocity $\omega/k_\para \sim \VAe$. For moderately high beta plasmas ($\beta_e \gtrsim 1$), the phase velocity is always smaller than the thermal velocity of the background Maxwellian electrons (i.e., $\omega/k_\para \lesssim \vth{e}$). In this case, the slope of VDF at the phase velocity depends only weakly on the electron temperature, leading to a weak dependence of Landau damping on $\beta_e$.

\section{Discussion}
\label{sec:discussion}

\subsection{Implications for Electron Acceleration}

It should be noted that the linear analysis discussed so far has been performed in the local plasma rest frame. However, concerning the scattering and acceleration of particles around the shock, the propagation direction of self-generated whistler waves relative to the shock is crucial. As we have seen, the \UAW instability favors high HTF \Alfven Mach number shocks. We have suggested that the critical condition may be roughly given by $\MAHTF \gtrsim 50$. It is well known that the so-called whistler critical Mach number $\MA^\mathrm{HTF*} = \sqrt{m_i/m_e}/2 \approx 21$ \cite{kennelQuarterCenturyCollisionless1985} is defined by the condition that a whistler wave emitted from the shock cannot propagate upstream beyond the critical value (This definition is for the phase velocity, but the one for the group velocity is similar). In other words, even though the whistler waves generated by the \UAW instability tend to propagate toward upstream in the plasma rest frame, they cannot escape away from the shock because the shock satisfying the instability condition is always super-critical with respect to the whistler critical Mach number. Therefore, the generated whistler waves will stay and accumulate within the shock transition layer, where they trigger the secondary instabilities.

Qualitatively, the idea of the self-generation of whistler waves by the shock-reflected electrons is compatible with the theory of SSDA. The critical assumption of SSDA is that the accelerated electrons are scattered in pitch angle to establish a nearly isotropic pitch-angle distribution within the shock transition layer. The present analysis indicates that the required isotropization may be a natural consequence of the wave generation triggered by the reflected electrons themselves. It is important to point out that the primary instabilities alone do not explain the full isotropization over the entire pitch angle because of the limited range of resonance in velocity space. The sequence of secondary instabilities found in this paper provides a reasonable scenario for the isotropization and the resulting electron confinement that plays the central role in particle acceleration.

In addition to the HTF Mach number dependence discussed above, several other parameters influence the wave generation. The upstream electron plasma beta, $\beta_e$, plays a particularly important role by controlling the number of electrons that can be reflected at the shock. A higher $\beta_e$ corresponds to a hotter upstream electron population, which increases the fraction of reflected electrons and thus the growth rate of the primary instabilities. The exact number of reflected electrons, however, is dependent on the profile and strength of the magnetic field and the cross-shock potential. Since the structure of the collisionless shock is determined by complicated nonlinear kinetic physics, we think the model provides only a qualitative prediction.

Our approach is complementary to fully kinetic PIC simulations. While PIC models capture self-consistent nonlinear dynamics, our framework enables a precise decomposition of resonance contributions, providing a theoretical baseline to interpret specific kinetic drivers. Notably, the whistler wave excitation conditions predicted by our linear analysis are consistent with the PIC simulation results (Riquelme \& Spitkovsky (2011)) in identifying oblique whistler waves in the shock transition layer, demonstrating that such waves are capable to pre accelerate electrons.

\subsection{Applications}

In-situ spacecraft observations at Earth's bow shock provide a natural test bed for the wave-generation mechanism discussed in this paper. A recent statistical analysis of MMS observations of the terrestrial bow shock has shown that the high-frequency whistler wave power (frequencies higher than $\sim 10\%$ of the local electron cyclotron frequency) within the shock transition layer is positively correlated with the HTF \Alfven\ Mach number \cite{amanoStatisticalAnalysisHighfrequency2024b}. Although the linear analysis cannot directly predict the saturated wave power, our model qualitatively explains the observed dependence. The same statistical study also suggested, albeit not conclusively, that shocks with higher upstream electron betas tend to exhibit stronger whistler activity within the shock transition layer. The electron beta dependence found in our analysis is also consistent with this observational trend. While we should keep in mind other possible mechanisms of whistler wave generation, such as the modified two-stream instability \cite{matsukiyoModifiedTwostreamInstability2003a,matsukiyoMicroinstabilitiesFootHigh2006}, our results indicate that the self-generation mechanism driven by shock-reflected electrons is likely to play a significant role in regulating the whistler wave activity within the shock transition layer.

The observational trend discussed above is also closely related to the onset of nonthermal electron acceleration via SSDA. The SSDA model predicts that efficient acceleration requires the whistler wave power in the shock transition layer to exceed a theoretical threshold \cite{amanoTheoryElectronInjection2022a,amanoObservationalEvidenceStochastic2020}. MMS statistics indicate that this condition is satisfied when $\MAHTF \gtrsim 30\text{--}60$, which is consistent with the observed efficient nonthermal electron acceleration \cite{okaWhistlerCriticalMach2006,laltiAdiabaticNonAdiabaticElectron2024}. The observed threshold in $\MAHTF$ is closely aligned with the condition suggested here: $\MAHTF \gtrsim 50$. Combining these, we suggest the scenario that the shock-reflected electrons drive whistler turbulence by themselves with sufficient power to sustain the SSDA once the HTF \Alfven\ Mach number exceeds the threshold.

Below the threshold, even though efficient electron acceleration is not expected, whistler waves may still be generated by the reflected electrons via the Landau resonance (e.g., $\tbn = 75 \degree$ case in Section \ref{sec:obliquity}). It is essentially the same mechanism as the one suggested for the so-called one-hertz whistlers frequently observed in the foreshock region upstream of planetary bow shocks \cite{sentmanObliqueWhistlerInstability1983a}. In this case, the instability growth rate is not sufficiently high and the shock can be sub-critical with respect to the whistler critical Mach number. Therefore, both the reflected electrons and the emitted whistler waves can escape upstream, consistent with the waves being observed in the upstream region. Therefore, we suggest that one-hertz whistlers are the outcome of the same wave generation scenario, but with the HTF \Alfven\ Mach number below the threshold.

The stringent condition required for triggering SSDA implies that efficient electron acceleration is relatively rare at Earth's bow shock, wich typically have moderate Mach numbers $\MA \sim 5\text{--}10$. By contrast, young SNR shocks generally have much higher Mach numbers, so that the HTF \Alfven Mach number can easily exceed the SSDA threshold with lower obliquities. In such systems, it is natural to expect that particle acceleration by SSDA is more common, which then provides a seed population for DSA. This explains the fact that young SNRs are often bright in nonthermal synchrotron emission from relativistic electrons. One possible issue is that the number of reflected electrons tends to decrease when the HTF \Alfven Mach numbers are excessively high (see Section \ref{sec:obliquity}), and the self-generation may become inefficient at very high Mach number shocks. We expect, however, that such shocks are accompanied by strong non-adiabatic electron heating and acceleration mechanisms associated with the Buneman instability and the ion-Weibel instability \cite{hoshinoNonthermalElectronsHigh2002,amanoElectronInjectionHigh2007,matsumotoElectronSurfingDrift2017,bohdanElectronPreaccelerationNonrelativistic2017,bohdanKineticSimulationsNonrelativistic2019,bohdanKineticSimulationsNonrelativistic2019a,jikeiSaturationLevelIon2024}, which are absent in moderate and low Mach number shocks. While electrostatic instabilities might also be present, they typically drive local diffusion along the magnetic field and are deferred to future study. These additional non-adiabatic electron energizations happening within the shock will enhance the reflection efficiency and, ultimately, the electron injection efficiency into DSA.

Another plausible application is shocks in the intracluster medium (ICM) of galaxy clusters. The presence of large-scale shock waves is inferred from observations of radio relics, which provide evidence for electron acceleration to relativistic energies \cite{vanweerenParticleAccelerationMegaparsec2010}. Since these shocks are thought to have relatively low Mach numbers and propagate in the high beta ICM plasma (i.e., different from the solar wind and the interstellar medium), dedicated kinetic simulations have been performed \cite{matsukiyoRELATIVISTICELECTRONSHOCK2011,kobzarElectronAccelerationRippled2021,guoNonthermalElectronAcceleration2014,guoNonthermalElectronAcceleration2014a}. The electron beta dependence found in our model suggests that efficient electron acceleration may occur at quasi-perpendicular ICM shocks even with lower Mach numbers.

Given possible applications to various shocks in the heliosphere and beyond, it is important to validate the predictions using fully kinetic PIC simulations. We emphasize that our discussion so far has assumed realistic values of the ion-to-electron mass ratio $m_i/m_e$ and the plasma-to-cyclotron frequency ratio $\wpe/\Wce$. In typical PIC simulations, however, artificial values of these parameters are often adopted to reduce computational cost. Since the electron beam velocity has to be faster than the whistler wave phase velocity, the theoretical threshold condition explicitly depends on the mass ratio \eqref{eq:ma_htf_cond}. Therefore, one must carefully choose the parameters to ensure that the instability condition is satisfied when an artificial mass ratio is used. Furthermore, the present analysis assumes non-relativistic electrons, which may not be strictly valid in most of the PIC simulations. Recall that the instability condition requires the typical shock-reflected electron beam speed to exceed the electron \Alfven speed, $\VAe = c (\Wce/\wpe)$. While $\VAe$ is non-relativistic in realistic, weakly magnetized plasmas such as the solar wind, it can become a significant fraction of the speed of light for commonly used values of $\wpe/\Wce$ in PIC simulations. It is therefore crucial to choose the simulation parameters so that the system remains in the appropriate regime to test the present model. Alternatively, the present analysis has to be extended to include relativistic effects for proper comparisons with PIC simulations. Indeed, there have been discussions on the generation of oblique whistler waves by shock-reflected electrons in PIC simulations, which were, however, all in the regime of relativistic electron beams \cite{matsukiyoRELATIVISTICELECTRONSHOCK2011,bohdanElectronForeshockHighMachnumber2022}.

\subsection{Validity of Assumptions}

The present analysis is based on a number of simplifying assumptions that allow us to capture the essential physics of the wave generation process while keeping the model tractable. First, the semi-analytical method is used with the cold-plasma dispersion relation to calculate the growth rate. Given that the electron temperature in the shock is finite and may not be negligible, this approach may not necessarily provide the exact quantitative values of the growth rate. Nevertheless, the qualitative identification of unstable modes should be possible since the wave growth is primarily determined by the slope of VDF at the resonance velocity. Therefore, the general trends and the relative stability properties discussed here are expected to remain valid even when a more realistic hot-plasma dispersion relation is used.

Second, the pitch-angle diffusion and the linear analysis are carried out in the local plasma rest frame, which is determined by assuming the frozen-in condition within the shock. However, we are aware that the plasma rest frame and the electron bulk velocity calculated in our model VDF differ quantitatively. The magnetic mirror force (decelerating the flow) and the cross-shock potential (accelerating the flow) both affect the electron bulk velocity within the shock. In particular, the electrons are strongly accelerated by the cross-shock potential in the deep shock transition layer, which may produce a substantial bulk velocity in the plasma rest frame. Theoretically, the potential will adjust by itself in such a way that the background flow and the electron bulk flow become nearly equal. This requires a self-consistent treatment of the potential and electron transport within the shock structure in the presence of scattering \cite{vanthieghemElectronHeatingHigh2024}. In contrast, the discrepancy appears in our model because we have chosen the potential as a free parameter. To avoid the problem, we have focused on the instability analysis near the upstream edge of the shock with large obliquity angles, where we have confirmed that the difference in flow velocities is smaller. Therefore, even though the issue remains a caveat of our model, we think the conclusions drawn here hold qualitatively.

Finally, we employ the Liouville mapping based on adiabatic theory and then apply the pitch-angle diffusion to mimic the non-adiabatic effect to construct the VDF. This is clearly not self-consistent, and the interpretation based on the model should be at best qualitative. Indeed, since the shock transition layer is an inhomogeneous and open system, the scattering should affect the spatial transport of particles, which is, however, completely ignored. While we have found a possibility that the self-generated primary instabilities may trigger the secondary instabilities and lead to isotropization, this requires that the pitch-angle diffusion time $D_{\mu\mu}^{-1}$ be sufficiently shorter than other competing time scales. For instance, one can easily anticipate that $D_{\mu\mu}^{-1}$ should be shorter than the ion gyroperiod, the dynamical time scale of the collisionless shock. Unfortunately, the linear analysis presented in this study does not provide any quantitative estimate of $D_{\mu\mu}$, which makes a more detailed discussion virtually impossible. Instead, we suggest that the condition $\gamma/\Wce \gtrsim 0.01$ should be satisfied to ensure wave growth within the shock transition layer. 
Nonlinear effects, such as pitch-angle scattering and wave-wave interactions, will ultimately determine the saturation level of the generated whistler turbulence and the resulting $D_{\mu\mu}$. A more comprehensive analysis is clearly needed to further test the applicability of the suggested idea in the future.
In particular, numerical tools capable of handling arbitrary VDFs, such as the Arbitrary Linear Plasma Solver (ALPS), could be utilized to validate the predicted growth rates for the non-Maxwellian distributions constructed in this study.

\section{Conclusion}
\label{sec:conclusion}

In this paper, we have investigated the instabilities driven by shock-reflected electrons at quasi-perpendicular shocks. We show that the reflected electrons may excite two primary instabilities on the whistler-mode branch: an upstream-directed anomalous-driven whistler (UAW) and a downstream-directed cyclotron-driven whistler (DCW). 
Both instabilities favor high \Alfven Mach numbers in the HTF, and the threshold is given roughly by $\MAHTF = \MA/\cos\tbn \gtrsim 50$ for the parameter regime relevant for Earth's bow shock. Under the condition, the generated waves will be convected back by the upstream flow and cannot escape from the shock. Consequently, the waves will accumulate within the shock transition layer.

Our analysis shows that as the primary modes (especially UAW) grow, they scatter the electrons and reshape the VDF, which subsequently triggers the secondary instabilities also on the whistler-mode branch (ULW, UCW). We suggest that this sequence of instabilities may efficiently scatter electrons over a broad range of pitch angles, substantially reducing the anisotropy. Although our analysis is based on linear theory, the identified instability sequence suggests a natural pathway toward sustained electron scattering.

The above conclusions are obtained under several simplifying assumptions. We assume a prescribed steady-state shock profile, adiabatic Liouville mapping, a parameterized cross-shock potential, and a simplified treatment of pitch-angle diffusion.
The linear growth rates are evaluated using a cold-plasma dispersion relation under the weak-growth approximation.
Nonlinear wave–particle feedback and shock non-stationarity are not included.
While these simplifications limit quantitative accuracy, they allow us to isolate the key resonance mechanisms responsible for wave generation.

These results imply that the reflected-electron-driven waves may contribute to their own confinement within the shock, which is necessary for electron acceleration by SSDA. Once particle acceleration is initiated, the energetic electron flux and the reflection efficiency will increase, which may then provide more free energy and enhance the wave intensity. This possible positive feedback between particle acceleration and wave generation may be the key to the self-confinement and the efficient electron acceleration by SSDA. A more comprehensive understanding of the feedback will ultimately lead to a solution for the electron injection problem.

\begin{acknowledgments}
This work was supported by JSPS KAKENHI grant No.~22K03697.
\end{acknowledgments}

\section*{Conflict of Interest Statement}
The authors have no conflicts to disclose.

\section*{Data Availability Statement}
Data sharing is not applicable to this paper as no new datasets were generated or analyzed during the current study.

\bibliographystyle{aipnum4-2}
\bibliography{reference}

\appendix
\section{Secondary Instabilities}
\label{appendix}

\begin{figure*}[p]
\centering

\begin{minipage}[t]{0.48\textwidth}
  \centering
  \includegraphics[width=\linewidth,height=0.85\textheight,keepaspectratio]{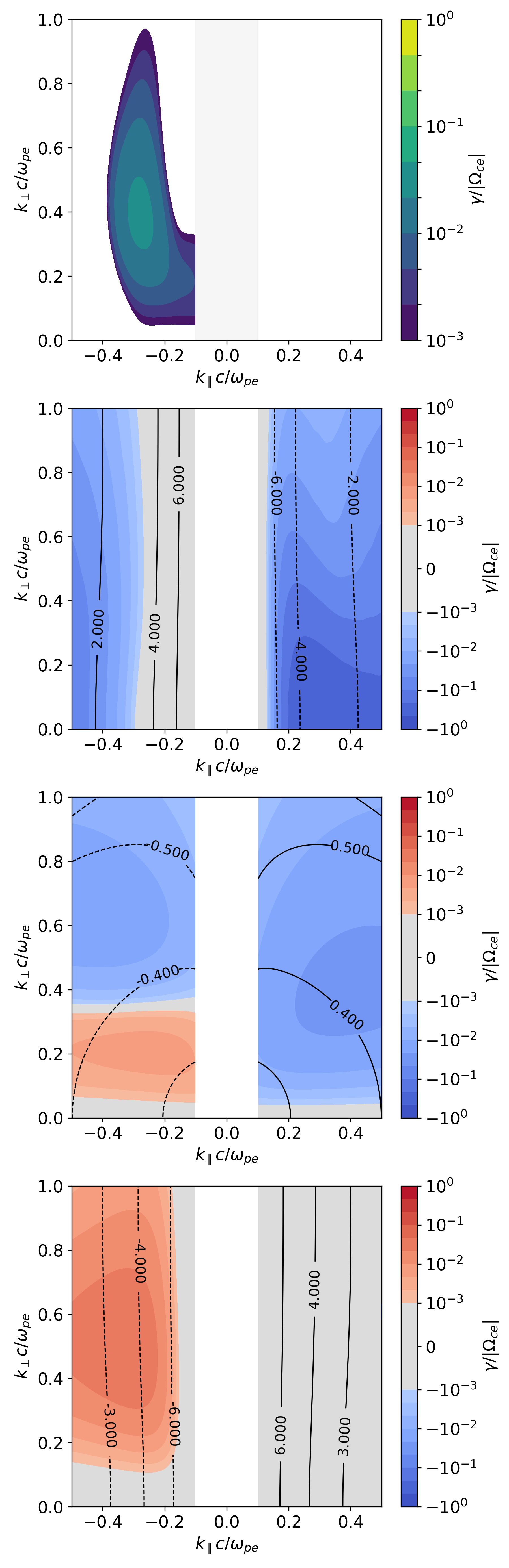}
  \caption{\label{fig:app1}
  Decomposition of the linear growth rate for the case of $D_{\mu\mu} t = 0.2$, corresponding to Fig.~\ref{fig:fig5}(d). The instability growth at lower wavenumbers is driven primarily by the Landau resonance ($n=0$). The format is the same as Fig.~\ref{fig:fig2}.}
\end{minipage}
\hfill
\begin{minipage}[t]{0.48\textwidth}
  \centering
  \includegraphics[width=\linewidth,height=0.85\textheight,keepaspectratio]{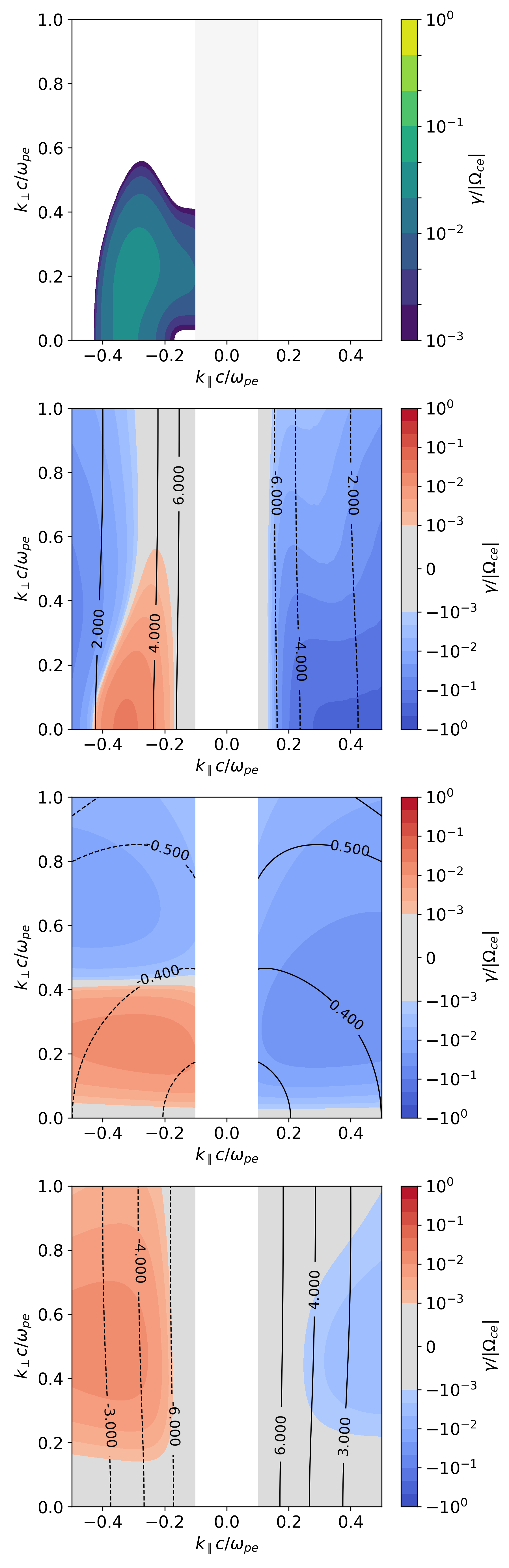}
  \caption{\label{fig:app2}
  Decomposition of the linear growth rate for the case of $D_{\mu\mu} t = 0.5$, corresponding to Fig.~\ref{fig:fig5}(f). The anti-parallel propagating instability is driven primarily by the normal cyclotron resonance ($n=-1$). The format is the same as Fig.~\ref{fig:fig2}.}
\end{minipage}
\end{figure*}

This appendix analyzes the mechanisms driving the secondary instabilities under the effect of pitch-angle diffusion. We first examine the cacse of $D_{\mu\mu} t = 0.2$, corresponding to Fig.~\ref{fig:fig5}(d). At this stage, diffusion transports electrons toward smaller parallel velocities.
Figure~\ref{fig:app1} presents the detailed linear growth rate map. Although the primary whistler instability remains driven by the anomalous cyclotron resonance ($n=+1$) at higher wavenumbers, the unstable region extends toward lower wavenumbers ($k_{\parallel} c/\omega_{pe} \simeq -0.2$, $k_{\perp} c/\omega_{pe} \simeq 0.2$). As shown in the third panel of Fig.~\ref{fig:app1}, this lower-wavenumber portion of the instability is driven by the Landau resonance contribution ($n=0$).

As diffusion proceeds further to $D_{\mu\mu} t = 0.5$, corresponding to Fig.~\ref{fig:fig5}(f), electrons cross into the positive parallel velocity region ($v_\para>0$), creating a secondary loss-cone feature. Figure~\ref{fig:app2} shows the growth rate decomposition for this case. A positive growth rate appears for anti-parallel propagation $k_{\para} c/\wpe \simeq -0.3$, $k_{\perp} c/\wpe \simeq 0$. The second panel confirms that this mode is driven by the normal cyclotron resonance ($n = -1$), as electrons in the positive $v_{\parallel}$ region resonate with anti-parallel waves. Meanwhile, contributions from anomalous cyclotron ($n = +1$) and Landau ($n = 0$) resonances remain positive.

\end{document}